\documentclass[twocolumn]{aastex631}

\usepackage[normalem]{ulem}
\usepackage{CJKutf8}

\shorttitle{Synchrotron Model of GRB Spectra}

\shortauthors{Wang et al.}

\begin{document}

\begin{CJK*}{UTF8}{gbsn}
\title{A Comprehensive Consistency Check between Synchrotron radiation and the Observed Gamma-ray Burst Spectra}

\correspondingauthor{Zhao-Yang Peng, Xiao-Hong Zhao,Bin-Bin Zhang}
\email{pengzhaoyang412@163.com,zhaoxh@ynao.ac.cn,bbzhang@nju.edu.cn}

\author{Dao-Zhou Wang}
\affiliation{College of Physics and Electronics, Yunnan Normal University, Kunming 650500, China \\}

\author[0000-0003-3659-4800]{Xiao-Hong Zhao}
\affiliation{Yunnan Observatories, Chinese Academy of Sciences, Kunming, China \\}
\affiliation{Center for Astronomical Mega-Science, Chinese Academy of Sciences, Beijing, China \\}

\author[0000-0001-6869-2996]{Zhao Joseph Zhang (张钊)}
\affiliation{School of Astronomy and Space Science, Nanjing
University, Nanjing 210093, China}

\author[0000-0003-4111-5958]{Bin-Bin Zhang}
\affiliation{School of Astronomy and Space Science, Nanjing
University, Nanjing 210093, China}
\affiliation{Key Laboratory of Modern Astronomy and Astrophysics (Nanjing University), Ministry of Education, China}
\affiliation{Department of Physics and Astronomy, University of Nevada Las Vegas, NV 89154, USA}

%\affiliation{AAS Journals Associate Editor-in-Chief}
\author{Zhao-Yang Peng}
\affiliation{College of Physics and Electronics, Yunnan Normal University, Kunming 650500, China \\}

\begin{abstract}

We performed a time-resolved spectral analysis of 53 bright gamma-ray bursts (GRBs) observed by \textit{Fermi}/GBM. Our sample consists of 1117 individual spectra extracted from the finest time slices in each GRB. We fitted them with the synchrotron radiation model by considering the electron distributions in five different cases: mono-energetic, single power-law, Maxwellian, traditional fast cooling, and broken power-law. Our results were further qualified through Bayesian Information Criterion (BIC) by comparing with the fit by empirical models, namely the so-called Band function and cut-off power-law models. Our study showed that the synchrotron models, except for the fast-cooling case, can successfully fit most observed spectra, with the single power-law case being the most preferred. We also found that the electron distribution indices for the single power-law synchrotron fit in more than half of our spectra exhibits flux-tracking behavior, i.e., the index increases/decreases with the flux increasing/decreasing, implying that the distribution of the radiating electrons is increasingly narrower with time before the flux peaks and becomes more spreading afterward. Our results indicate that the synchrotron radiation is still feasible as a radiation mechanism of the GRB prompt emission phase.

\end{abstract}

\keywords{gamma-ray burst: general --- radiation mechanisms: non-thermal --- methods: spectral fitting}

\section{Introduction} \label{sec:intro}
\end{CJK*}

The radiation mechanism of the sub-MeV emission in gamma-ray bursts (GRBs) is still an unsolved issue, despite extensively studied in the past decades. One helpful tool to confront observed data to physical radiation models is the spectral fit. It has been widely accepted that GRB spectra can be modeled using some empirical models, such as the Band function (BAND, Band et al. 1993), cut-off power-law (CPL), or smoothly broken power-law models (SBPL, Preece et al. 2000; Kaneko et al. 2006; Gruber et al. 2014). Those models, however, are purely mathematical functions and possess no direct physical meanings. From the theoretical point of view, the observed GRB photons can be generated either through thermal or non-thermal radiation, both directions having been investigated with intensive efforts by, e.g., relating the thermal-like spectra to the photospherical radiation of the fireball (M\'esz\'aros \& Rees 2000; M\'{e}sz\'{a}ros et al. 2002; Pe'er et al. 2006) and non-thermal spectra to synchrotron radiation of Poynting flux in a large emission radius or internal shocks (e.g., Zhang et al. 2016). However, the measured low-energy photon indices through the BAND fit to the observed spectra peak at around $\sim -1$ (Preece et al. 2000; Kaneko et al. 2006; Poolakkil et al. 2021), which is inconsistent with any segment of the theoretical synchrotron spectrum and also deviates from the photosphere model.

In the internal shock model, the expected low-energy photon slope is $\sim -3/2$, different from the observed $-1$ (Poolakkil et al. 2021). This is so-called fast cooling problem (Ghisellini et al. 2000). On the other hand, the ``death-line" limit of the low-energy index of the synchrotron model is $-2/3$, which has been violated by some GRBs (Crider et al. 1998; Preece et al. 1998, 2002). Mounting efforts have been made in order to solve these problems. For example, additional physical effects such as Inverse Compton (IC) cooling in Klein-Nishina (KN) regime is considered in Derishev et al. (2001); Daigne et al. (2011); Wang et al. (2009); Nakar et al. (2009). A decaying magnetic field in the emission region is introduced in Pe'er \& Zhang (2006); Zhao et al. (2014); Uhm \& Zhang (2014); Geng et al. (2018). Liu et al. (2020, 2021) modified the injection rate of electrons from a usually-used constant form to a flexible time-dependent function to phenomenally convey the observed spectral shapes. These models all resulted in flatter electron distributions than that in the fast-cooling case and better explained the observed data. Recently, Burgess et al. 2020 claimed that a synchrotron model with a flexible synchrotron cooling break can accommodate the observed spectra of bursts from a large sample and that the model even overcomes the synchrotron ``death-line" problem. In all the above models, the electron distribution is a crucial factor that determines the shape of photon spectrum and diagnoses the physics of particle acceleration and radiation.

In this paper, we first performed an extensive study on nearly all possible forms of electron distributions and derived the consequent synchrotron spectra models (\S 2). Then, by processing a large sample of brightest GRBs from \textit{Fermi} Gamma-ray Burst Monitor (GBM) data (\S 3), we fit the observed time-resolved spectra with our models, present the model comparisons, analyze the parameter distributions and their correlations of the best models (\S 4). Finally, we summarize and discuss the implications of our results in \S 5.

\section{The Models}

We consider seven different models including five synchrotron models from various electron distributions and two empirical models, as listed below.

\begin{enumerate}
 \item Synchrotron model with a power-law electron distribution (SYNPL). The electron energy distribution follows:
  \begin{equation}
 \frac{dn_{e}(\gamma_{e})}{d\gamma_e} \propto \gamma _{e}^{-p}
 \label{ele}
\end{equation}
 where $p$ is the power-law index.

The observed flux density of the synchrotron radiation, $ F_{\nu }\ \mathrm{\left [erg\ s^{-1}\ cm^{-2}\ Hz^{-1} \right ]} $ can be calculated as
\begin{eqnarray}
F_{\nu }&=& C\int_{\gamma _{m}}^{\gamma _{max}}\frac{dn_{e}(\gamma_{e})}{d\gamma_e}P(\nu,\gamma _{e})d\gamma _{e},
 \label{intensity}
\end{eqnarray}
where $C$ is a normalization factor and $\gamma_{m}$ and $\gamma_{max}$ are the minimum and maximum injection energies of the electron population, respectively. $\gamma_{max}$ is fixed to $10^7$ in our analysis. The synchrotron emissivity averaged over an isotropic distribution of pitch angles scales, $P(\nu,\gamma_{e})$, is given as (Crusius \& Schlickeiser 1986):
\begin{equation}
P(\nu,\gamma_{e})\propto t^{2}\left \{ K_{4/3}(t)K_{1/3}(t)-\frac{3}{5}t\left [ K_{4/3}^{2}(t)-K_{1/3}^{2}(t) \right ] \right \},
\label{single_ele_Pnu}
\end{equation}
where $t = \nu /(3\gamma _{e}^{2}\nu _{L}), \nu _{L} = eB/(2\pi m_{e}c)$, $\gamma_e$ is the electron energy, $\nu$ is the synchrotron photon frequency, $B$ is the magnetic field and $ K_{n}(t)$ is the modified Bessel function of order $n$. The magnetic field is fixed to be $B=5000$ G throughout this paper, since it does not change the shape of synchrotron spectrum.

For any given free parameter set [$C$, $\gamma_m$, $p$], one can calculate the observed flux via Eq. (2), which can be further compared with observed data.

 \item Synchrotron model with a broken power-law electron distribution (SYNBPL). The electron energy distribution follows:

 \begin{equation}
 \frac{dn_{e}(\gamma_{e})}{d\gamma_e} \propto\left\{\begin{array}{clcc}
\gamma _{e}^{-p_{1}}, \gamma _{m}<\gamma _{e}<\gamma _{b}, \\
\gamma _{b}^{p_{2}-1}\gamma _{e}^{-p_{2}}, \gamma _{b}< \gamma _{e} < \gamma _{max},
\end{array}\right.
\label{ele}
\end{equation}
where $\gamma_b$ is the break energy of electrons. The index, $p_1$, is set as a free parameter in our model, which can flexibly account for the effects of several cases such as IC cooling in Klein-Nishina (KN) regime (e.g. Wang et al. 2009; Nakar et al. 2009), magnetic field decay (e.g., Zhao et al. 2014; Uhm \& Zhang 2014), or a time-dependent injection rate (Liu et al. 2020). Note that $p_1=2$ corresponds to the fast cooling case, while $p_1=p_2-1$ corresponds to the slow cooling case. The SYNBPL model actually includes the SYNPL case, provided $p_1<1/3$ or $p_2\gg p_1$. Note that in the traditional fast cooling model of GRBs in the prompt phase, the magnetic field in GRB emission region is as high as the equipartition field so that the 1/3 section should be out of the gamma-ray bands toward low energies. However, some recent studies, e.g., Burgess et al. (2020), show that the 1/3 section can fall into gamma-ray bands, in the case that the magnetic field is much lower than the equipartition field. Our SYNBPL model also includes the case discussed in Burgess et al. (2020) through the configuration of $\gamma_b$ or $\gamma_m$.

Similar to case 1, for any given free parameter set [$C$, $\gamma_m$, $\gamma_b$, $p_1$, $p_2$], one can substitute Eq. (4) to Eq. (2) to derive the observed flux.

\item Synchrotron model with the fast cooling electron distribution (SYNFC). The distribution is:
 \begin{equation}
\frac{dn_{e}(\gamma_{e})}{d\gamma_e}\propto\left\{\begin{array}{clcc}
\gamma_e^{-2}, &\gamma _{e}<\gamma_{m}\\
\gamma_e^{-(p+1)}, & \textrm{otherwise},
\end{array}\right.
\label{ele}
\end{equation}
For any given free parameter set [$C$, $\gamma_m$, $p$], one can substitute Eq. (5) to Eq. (2) to derive the observed flux.

\begin{figure*}
\flushleft
\vspace{-0.1cm}
\hspace*{2cm}
\vspace*{-1.0cm}
 \includegraphics[width=7cm,height=5cm]{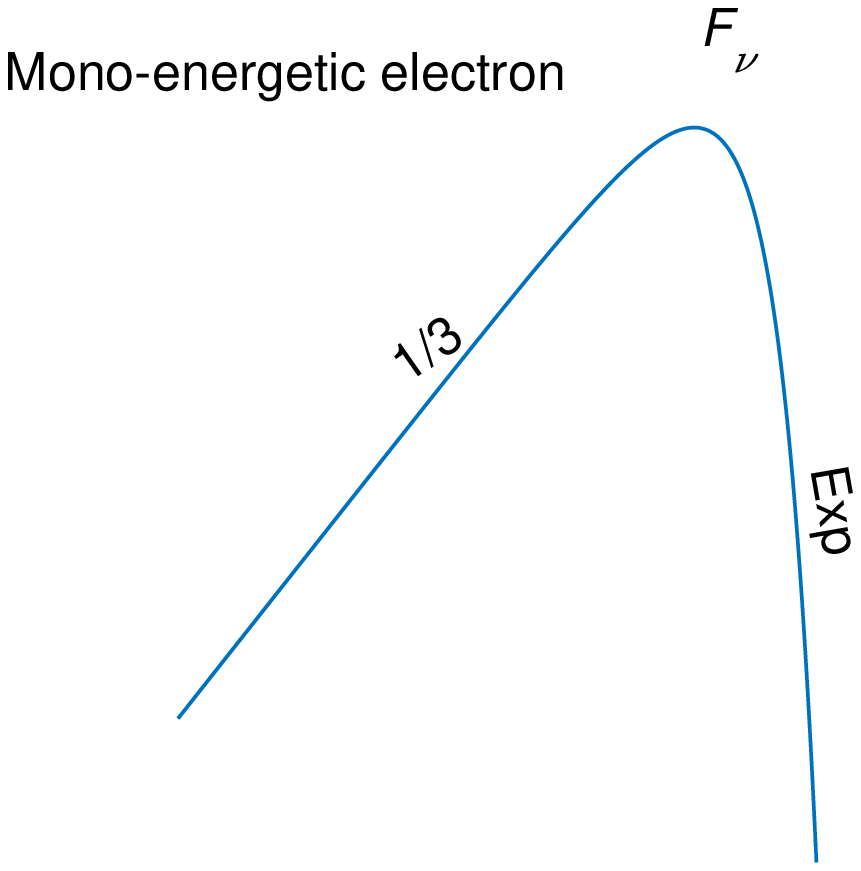}
 \includegraphics[width=7cm,height=5cm]{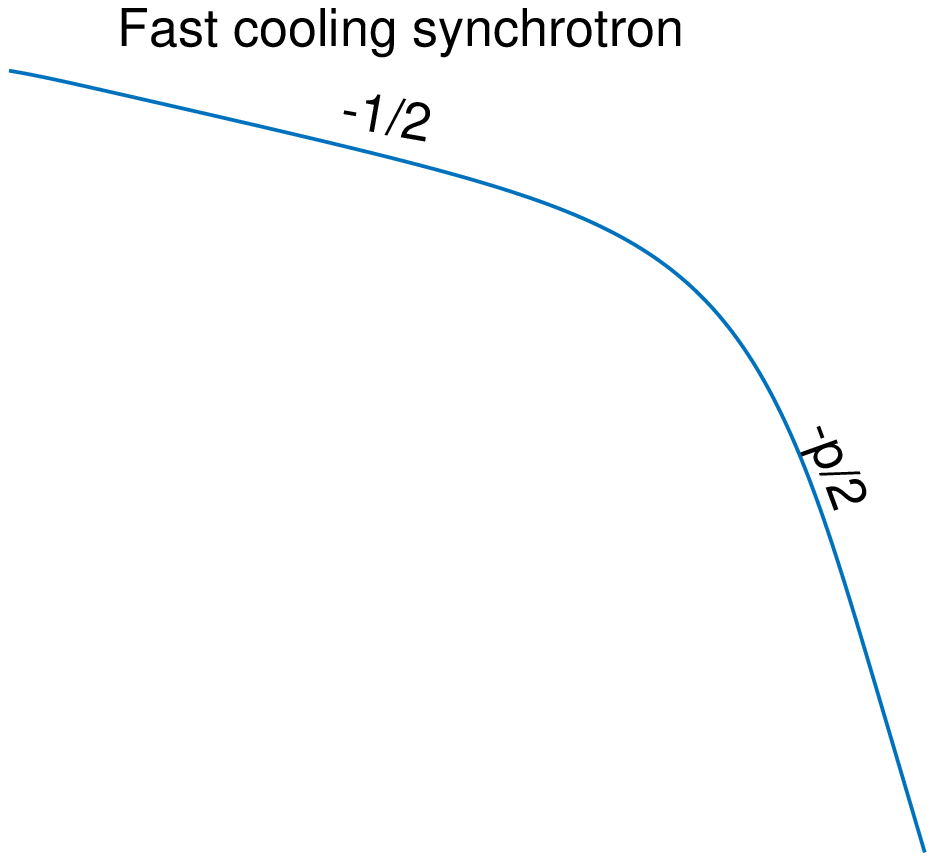}
\hspace*{2cm}
\vspace{-1.0cm}
 \includegraphics[width=7cm,height=5cm]{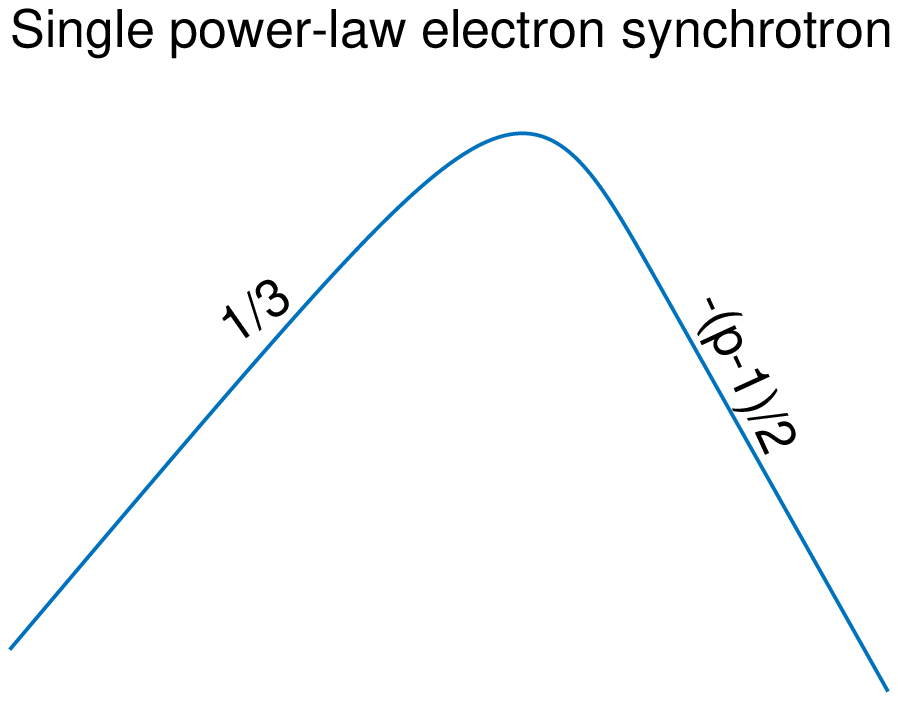}
 \includegraphics[width=7cm,height=5cm]{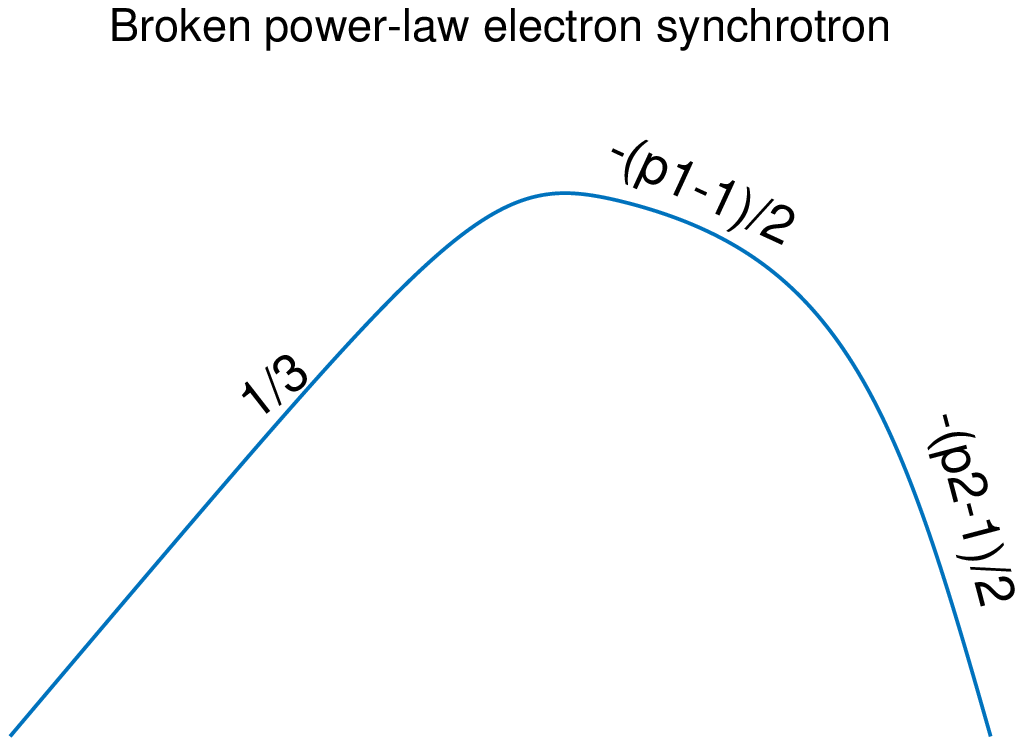}
\hspace*{2cm}
\vspace{-0.5cm}
 \includegraphics[width=7cm,height=5cm]{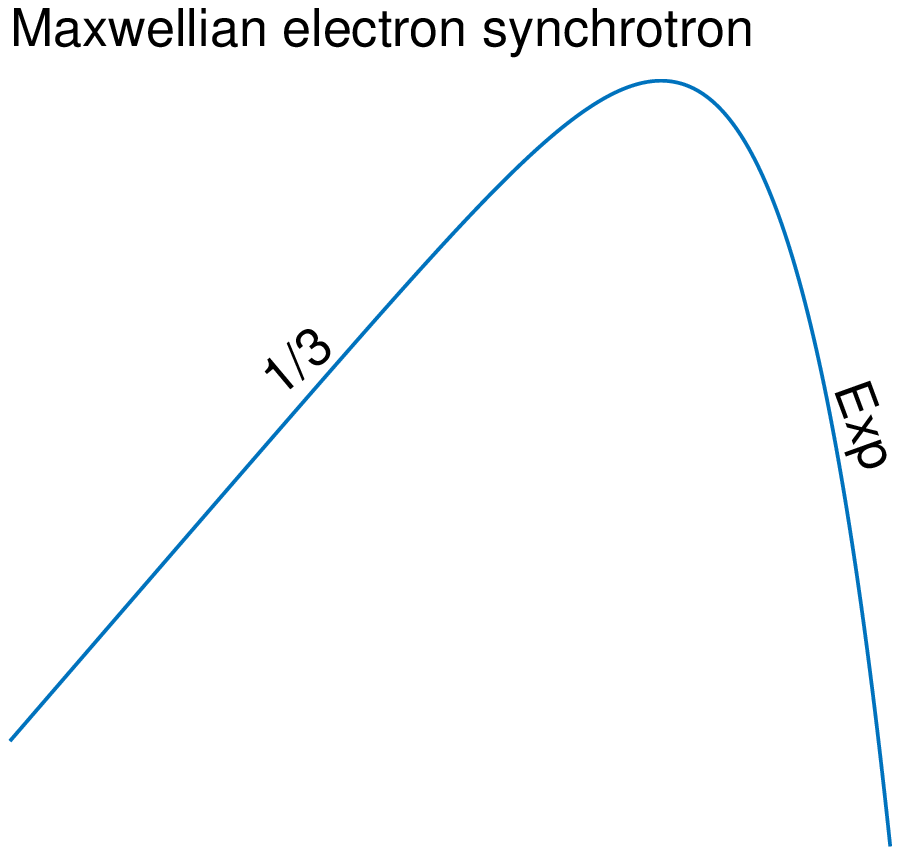}
\hspace*{2cm}
\vspace{1cm}
 \caption{Schematic sketch of the $F_{\nu}$ synchrotron spectra from various electron distributions, including mono-energetic, fast-cooling, single power-law, broken power-law and Maxwellian.}
 \label{Fv}
\end{figure*}

\item Synchrotron model with a Maxwellian electron distribution (SYNMAX). The distribution is
\begin{equation}
n_{e}(\gamma_{e} )\propto \left (\frac{\gamma _{e}}{\gamma _{th}} \right)^{2} \mathrm{exp}\left (-\frac{\gamma _{e}}{\gamma _{th}} \right)
\label{MAX_ele_Pnu}
\end{equation}
where $\gamma _{th}$ is the dimensionless electron temperature. Such a distribution
has been found in a long-term particle-in-cell simulation of an unmagnetized shock and even carries most of the shock energy (Spitkovsky 2008; Giannios \& Spitkovsky 2009).

For any given free parameter set [$C$, $\gamma_{th}$], one can substitute Eq. (6) to Eq. (2) to derive the observed flux.

\item Synchrotron model with mono-energetic electron distributions (SYNSE). In this model, the energy of all the electrons are assumed to be the same, following a distribution of

 \begin{equation}
\frac{dn_{e}(\gamma_{e})}{d\gamma_e}\propto\left\{\begin{array}{clcc}
C, &\gamma _{e}=\gamma_{m}\\
0, & \textrm{otherwise},

\end{array}\right.
\label{ele}
\end{equation}
where $C$ is a normalization factor and $\gamma_m$ is the Lorentz factor of all the electrons.

For any given free parameter set [$C$, $\gamma_{m}$], one can substitute Eq. (7) to Eq. (2) to derive the observed flux.

 \item Band function model (BAND). The widely-used empirical model introduced by Band et al. 1993 follows
 \begin{tiny}
 \begin{equation}
f_{\textrm{Band}}(E)=A\left\{\begin{array}{clcc}
(\frac{E}{100\mathrm{keV}})^{\alpha }\mathrm{exp}\left[-\frac{(\alpha+2)E}{E_{p}} \right ], E<E_{c}, \\
(\frac{E}{100\mathrm{keV}})^{\beta }\mathrm{exp}(\beta -\alpha )(\frac{E_{c}}{100\mathrm{keV}})^{\alpha-\beta }, E\geq E_{c}
\end{array}\right.
\end{equation}
 \end{tiny}
where $A$ is the normalization factor at 100 keV in units of $ \mathrm{ph\ s^{-1}\ cm^{-2}\ keV^{-1}}, \alpha$ and $\beta$ are the low- and high-energy power-law photon indices, respectively, $E_{p}$ is the peak energy in units of keV, and $E_{c} (E_{c}=\left ( \frac{\alpha -\beta }{\alpha +2} \right )E_{p})$ is the characteristic energy in units of keV.

One can calculate the flux giving any free parameter set of [$A$,$\alpha$,$\beta$,$E_p$].

\item Cut-off Power Law model (CPL). Another empirical model follows
\begin{equation}
f_{\textrm{CPL}}(E)=A(\frac{E}{100\textrm{keV}})^{\alpha }\textrm{exp}\left [ -\frac{(\alpha+2)E}{E_{p}} \right ],
\end{equation}
where $A$ is the normalization factor at 100 keV in units of $ \mathrm{ph\ s^{-1}\ cm^{-2}\ keV^{-1}}, \alpha$ is the low-energy power-law photon index, $E_{p}$ is the peak energy in units of keV.

One can calculate the flux giving any free parameter set of [$A$,$\alpha$,$E_p$].

\end{enumerate}

Figure \ref{Fv} depicts the synchrotron spectrum ($F_{\nu}$) with the models 1-5. In Section 3, we will employ those models as well as the two empirical models (6-7) to fit the observed spectra of a large sample of bright GRBs.

\begin{table*}
\renewcommand\tabcolsep{2.0pt}
\renewcommand\arraystretch{1.2}
 \centering
 \caption{GRBs, detectors, time slices and the number of spectra selected in the data fitting.}
 \begin{tabular}{cccccccc}
 \hline\hline
 GRB & Detectors &$\Delta T_{source}$ & Spectral number &GRB & Detectors &$\Delta T_{source}$ & Spectral number\\
 \hline
 130121835 & n4 n8 b1 & 0.00-13.00 & 15 & 160215773 & n4 n5 b0 & 168.00-207.87 & 18 \\
 130305486 & n6 n9 b1 & 2.55-14.46 & 12 & 160530667 & n1 n2 b0 & 4.25-9.36 & 14 \\
 130327350 & n0 n1 b0 & 0.00-34.82 & 12 & 160720767 & n2 n5 b0 & -1.02-100.41 & 66 \\
 130502327 & n6 n7 b1 &5.75-8.35; 9.65-32.24 & 20 & 160905471 & n6 n9 b1 & 7.15-23.79 & 22 \\
 130518580 & n3 n4 b0 & 20.25-32.97 & 13 & 160910722 & n1 n5 b0 & 7.55-14.22 & 13 \\
 130606497 & n8 nb b1 & -1.02-65.23 & 29 & 170114917 & n1 n2 b0 & 0.00-11.77 & 7 \\
 130609902 & n4 n5 b0 & 5.85-25.72 & 17 & 170115743 & n0 n1 b0 & -1.02-1.05; 1.05-43.46 & 7 \\
 130614997 & n0 n3 b0 & 0.23-4.65 & 5 & 170210116 & n2 na b0 & -6.14-100.48 & 19 \\
 130815660 & n3 n4 b0 & 30.00-44.03 & 9 & 170308221 & n9 na b1 & 7.45-13.85 & 10 \\
 130821674 & n6 n9 b1 & -21.05-100.5 & 58 & 170409112 & n2 n5 b0 & 25.28-67.35 & 32 \\
 130925173 & n6 n7 b1 & 25.05-167.85 & 38 & 170522657 & n1 n2 b0 & 0.00-5.68 & 6 \\
 131014215 & n9 nb b1 & 1.33-3.97 & 23 & 170527480 & n3 n6 b0 & -1.02-37.49 & 23 \\
 131108862 & n0 n3 b0 & -1.02-18.05 & 16 & 170826819 & na nb b1 & 2.15-10.22 & 11 \\
 131127592 & n2 n5 b0 & 0.65-7.34; 8.37-9.40; 16.62-17.65
 & 9 & 170906030 & n3 n4 b0 & -1.02-124.14 & 26 \\
 131231198 & n0 n3 b0 & 18.65-39.45 & 30 & 171210493 & n0 n1 b0 & 3.65-18.45 & 22 \\
 140206275 & n0 n1 b0 & 8.00-35.86 & 39 & 171227000 & n3 n5 b0 & 15.25-28.35 & 27 \\
 140306146 & n3 n4 b0 & -1.02-55.71 & 18 & 180113418 & n2 n9 b1 & 8.35-27.25 & 25 \\
 140723499 & n9 na b1 & 24.65-37.55 & 18 & 180210517 & n0 n1 b0 & 6.25-30.90 & 26 \\
 141028455 & n6 n9 b1 & 8.45-22.65 & 15 & 180305393 & n1 n2 b0 & 2.65-8.85 & 12 \\
 150105257 & n8 nb b1 & -4.10-80.45 & 36 & 180720598 & n6 n7 b1 & -0.06-60.22 & 66 \\
 150213001 & n7 n8 b1 & 0.37-3.69 & 7 & 190114873 & n3 n4 b0 & 0.00-6.00; 16.14-16.63 & 19 \\
 150306993 & n4 n5 b0 & 0.00-23.88 & 14 & 190530430 & n1 n2 b0 & 8.00-19.82 & 38 \\
 150314205 & n1 n9 b1 & 0.00-12.00 & 15 & 190531840 & n0 n3 b0 & 13.55-35.55 & 30 \\
 150510139 & n1 n5 b0 & 0.00-40.80 & 20 & 190731943 & n6 n9 b1 & 1.16-13.82 & 15 \\
 150902733 & n0 n3 b0 & 4.95-15.45 & 16 & 191227069 & n2 n5 b0 & 12.15-26.81 & 19 \\
 151021791 & n9 na b1 & 0.00-10.00 & 6 & 200412381 & n7 n8 b1 & 7.63-10.94 & 13 \\
 160113398 & n8 nb b1 & 29.65-45.90 & 21 & & & & \\ \hline
 \end{tabular}
\end{table*}

\section{Fit and Results} \label{sec:floats}
The Gamma-ray Monitor on board \textit{Fermi} has been triggered by more than three thousand GRBs to date (Poolakkil et al. 2021). We used the data observed by the GBM, which consists of twelve sodium iodide (NaI, 8 keV to 1 MeV) detectors and two bismuth germanium oxide (BGO, 200 keV to 40 MeV) detectors and covers an energy range from 8 keV to 40 MeV (Bissaldi et al. 2009; Meegan et al. 2009). In order to avoid the K-edge\footnote{ https://fermi.gsfc.nasa.gov/ssc/data/analysis/GBM caveats.html} at 33.17 keV in spectrum analysis, we use energy spectrum analysis range of 8 keV $\sim$ 30 keV and 40 keV $\sim$ 900 keV for NaI detectors, and use 250 keV $\sim$ 40 MeV for BGOs. We use the Time-Tagged Events (TTE) data with the highest temporal and spectral resolutions for spectrum analysis. Following Yu et al. (2016), we select 53 long bright GRBs (listed in Table 1) from Fermi Science Support Center (FSSC) according to the following criteria: the energy fluence $f$ and $\mathrm{peak\ photon\ flux}\ F_{p}$ in $\mathrm{10\ keV \sim 1\mathrm{\ MeV}}$ bands with the time bin size of 64, 256, and 1024 ms are greater than $4\times 10^{-5}\mathrm{erg\ cm^{-2}}$ and $20\mathrm{\ ph\ s^{-1}\ cm^{-2}}$, respectively. Data from the two brightest NaI detectors and the brightest BGO detector are selected. The three detectors used in each GRB, the spectrum extraction range, and the number of spectra are shown in Table 1. To study the spectral evolution within a GRB, we divide the whole burst into finest slices allowed by statistics using the same criteria as that in Zhang et al. (2018), i.e., each bust is split averagely within $T_{90}$. Different time-bin lengths from 0.11s to 6.16s are adopted for different bursts, depending on the brightness of the bursts. Furthermore, in each time slice of each burst, the averaged net counts per energy channel within the 128 energy channels are required to be larger than 20. For each slice, we extract source spectra, background spectra, and corresponding response files following the standard procedure described in Zhang et al. (2016, 2018). Our final sample consists of 1117 sets of spectra in those 53 bright GRBs. Each set of the spectra in each slice are then fitted by the aforementioned seven models using our own spectral fitting package McSpecFit (Zhang et al. 2016).

\begin{figure*}
\gridline{\fig{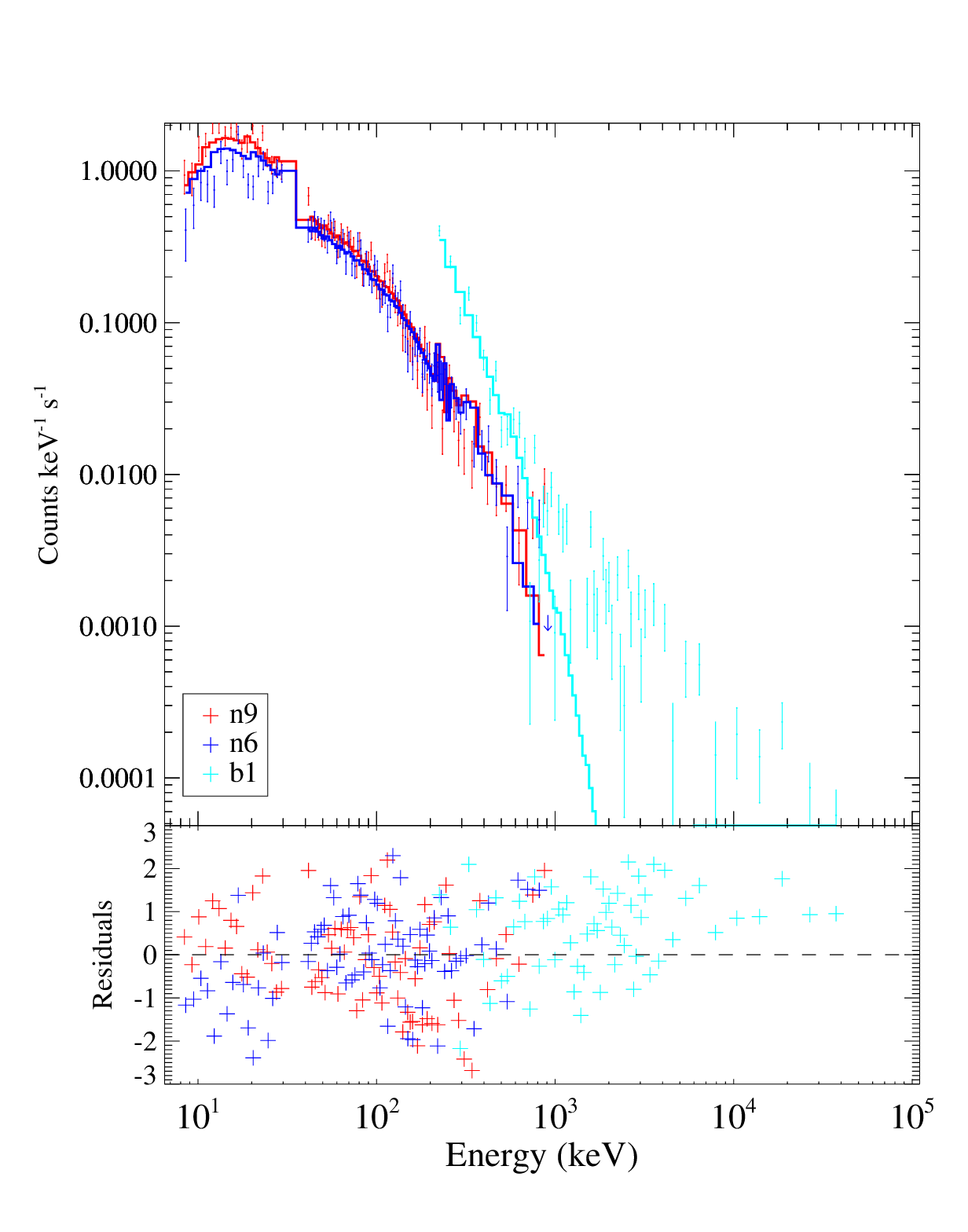}{0.3\textwidth}{(CPL)}
 \fig{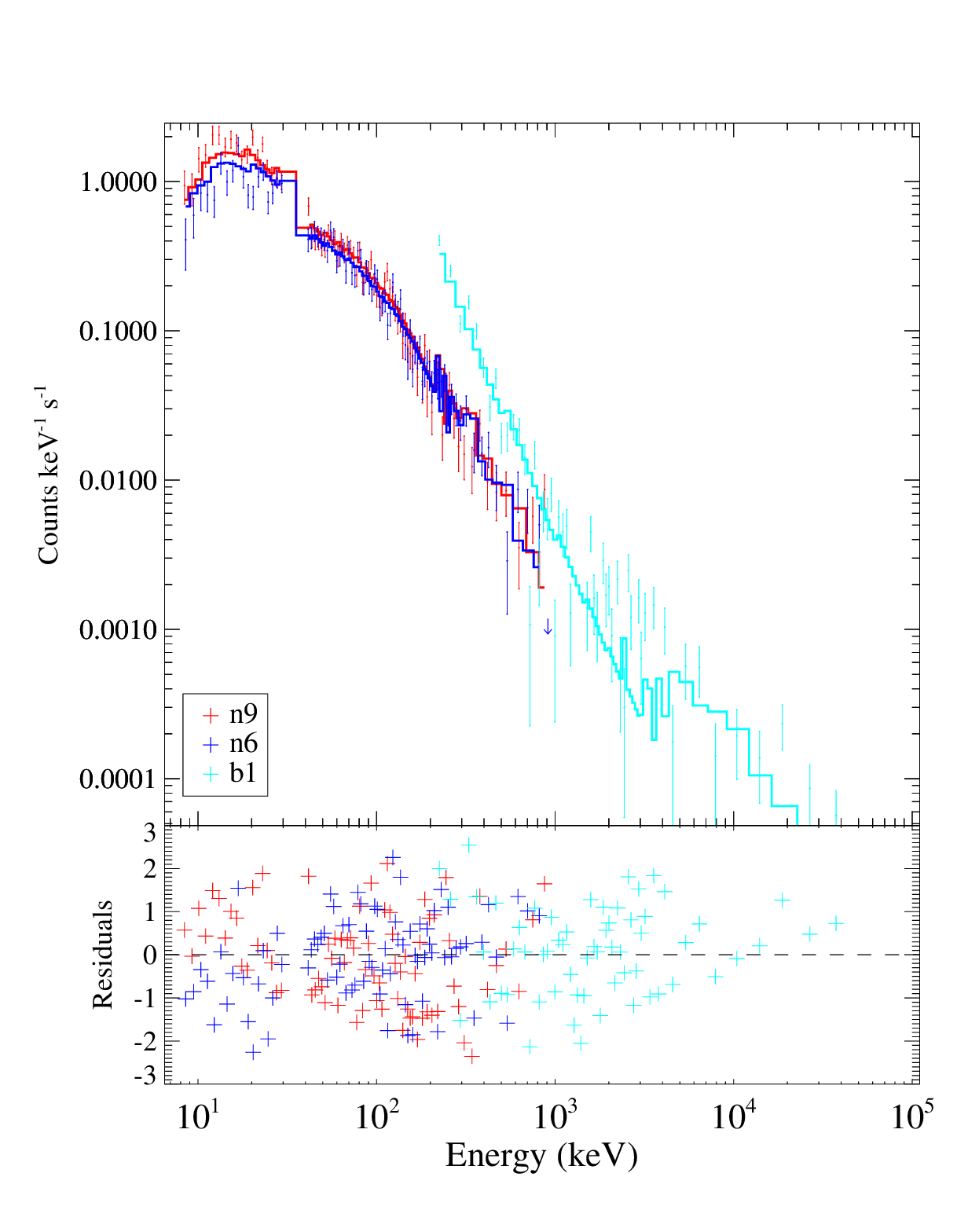}{0.3\textwidth}{(BAND)}
 \fig{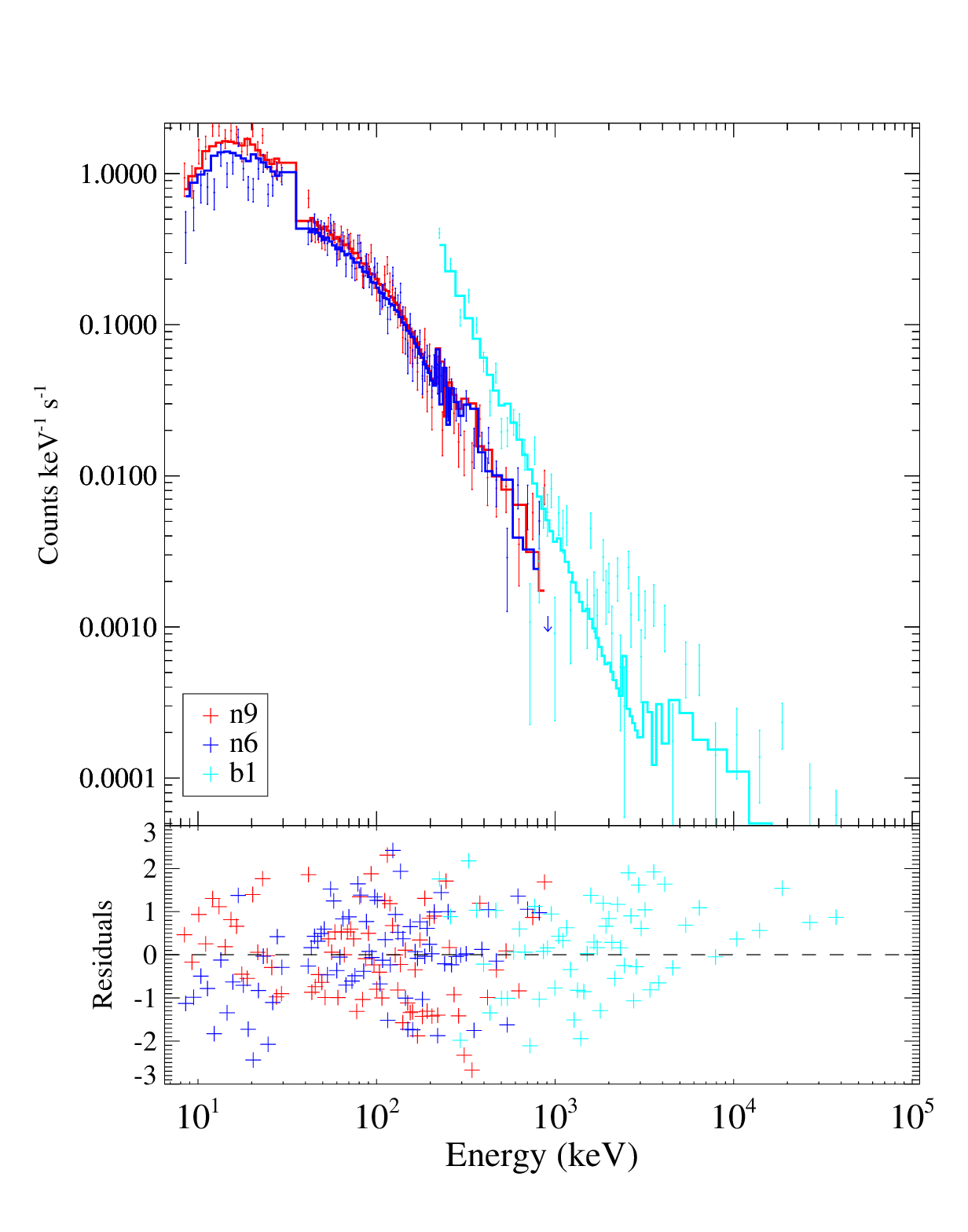}{0.3\textwidth}{(SYNPL)}}

\gridline{\fig{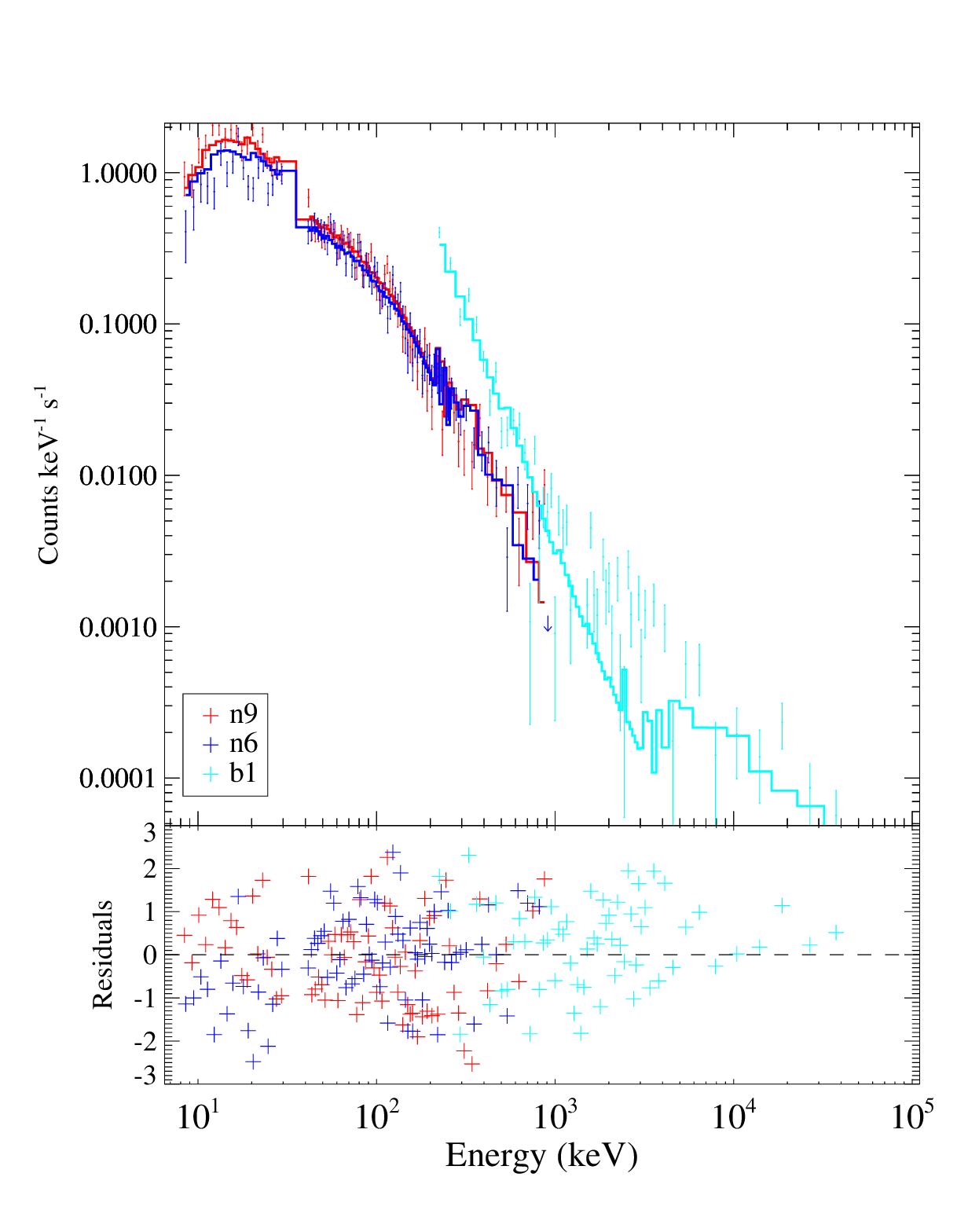}{0.3\textwidth}{(SYNBPL)}
 \fig{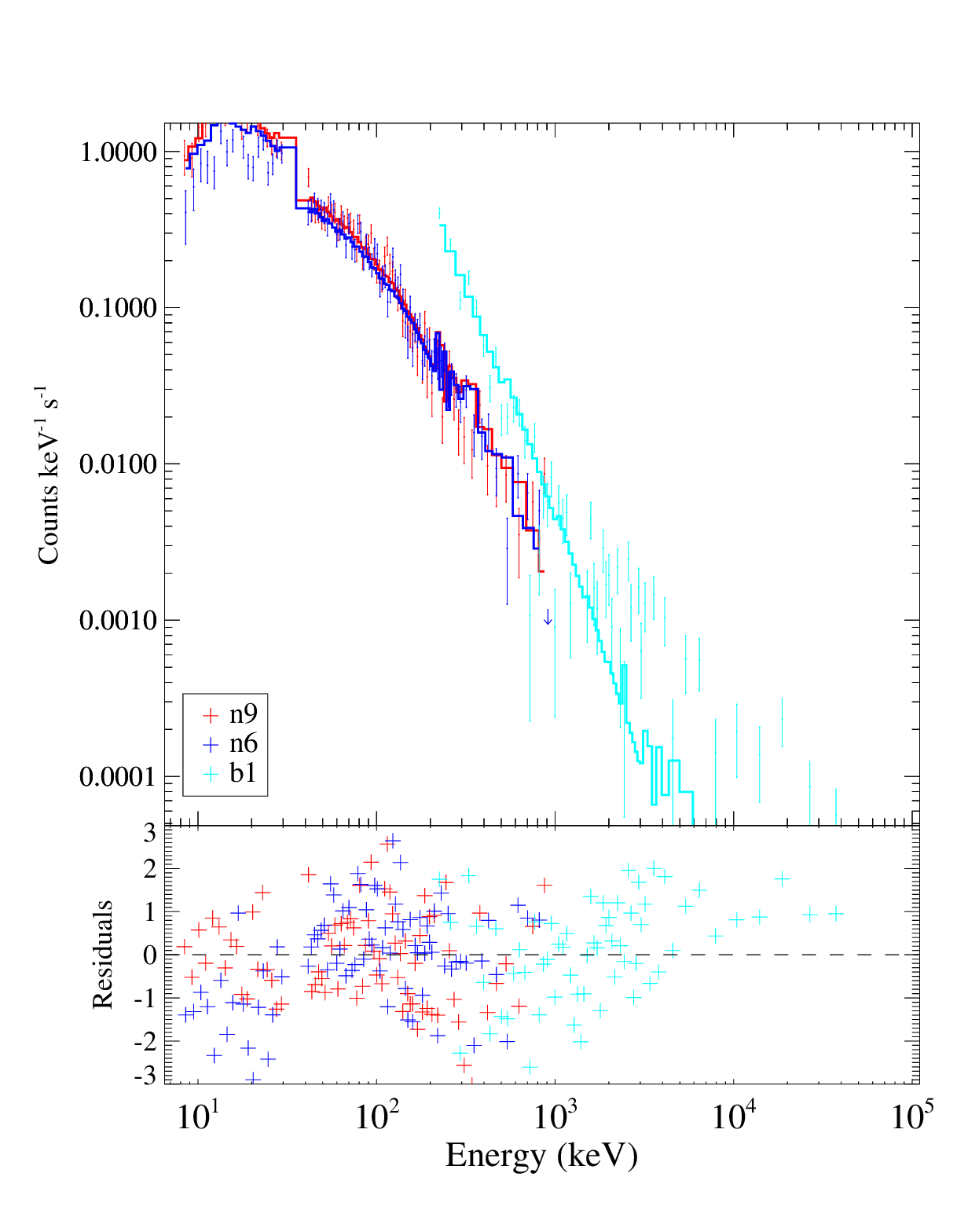}{0.3\textwidth}{(SYNMAX)}
 \fig{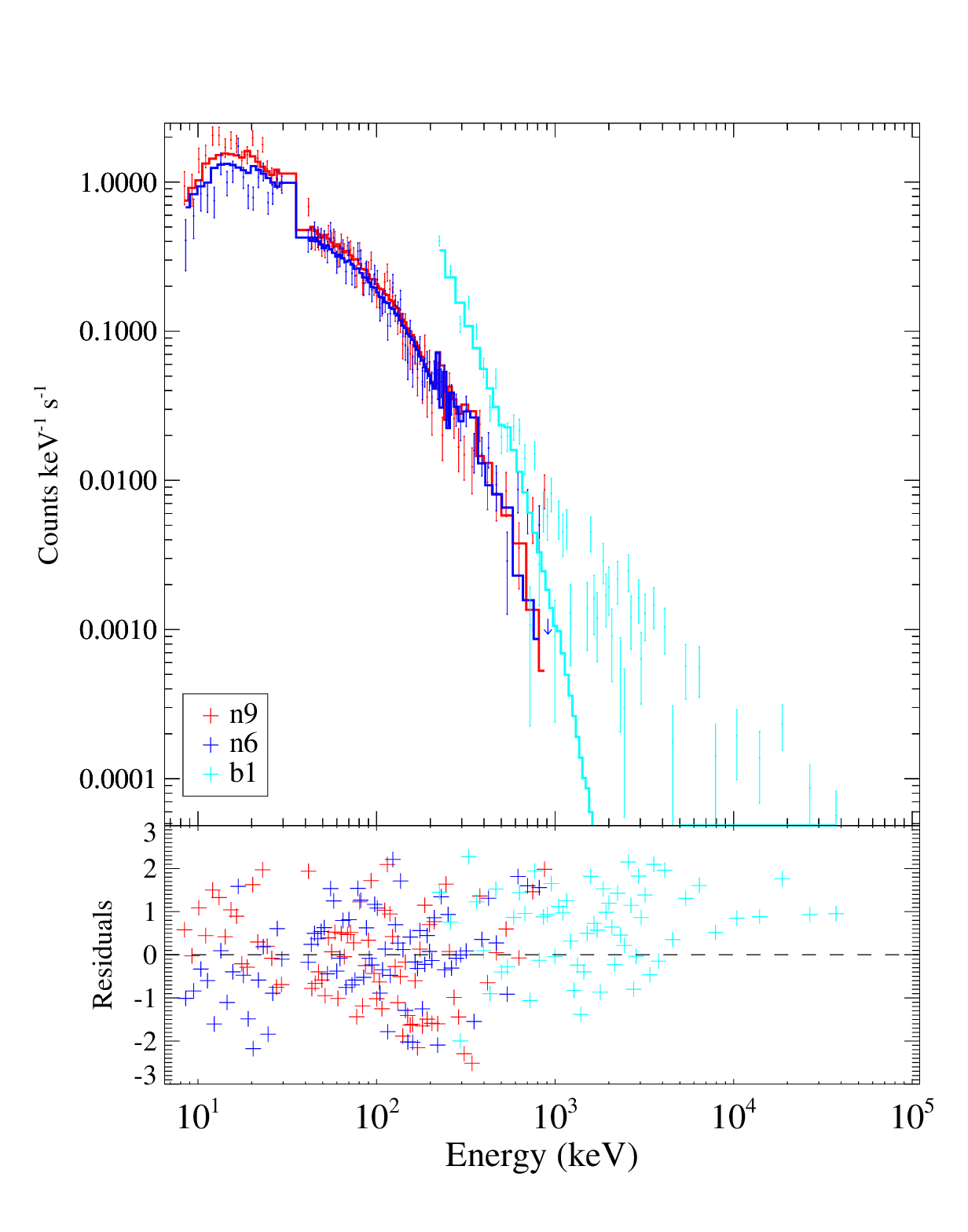}{0.3\textwidth}{(SYNSE)}}

\gridline{\fig{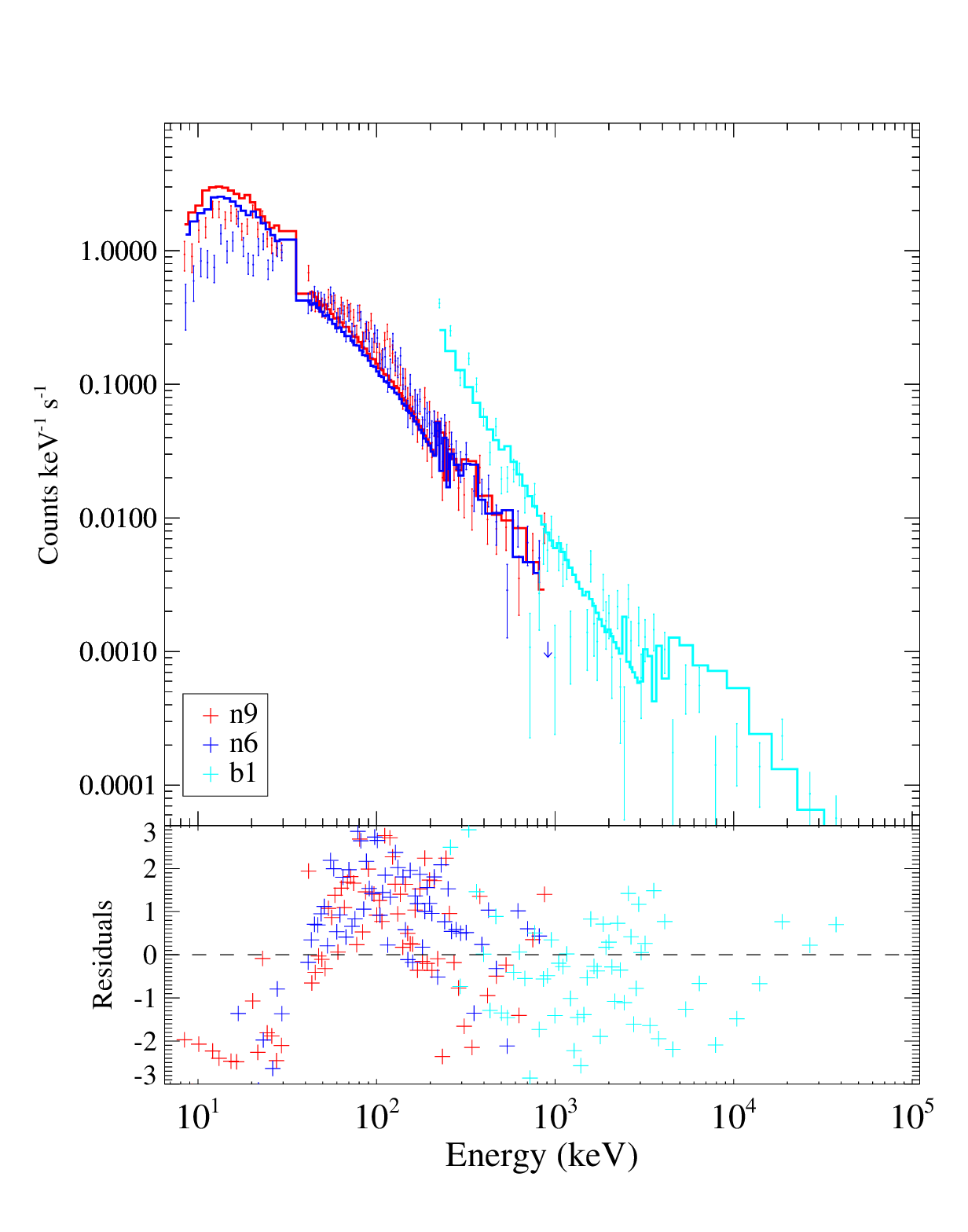}{0.3\textwidth}{(SYNFC)} \fig{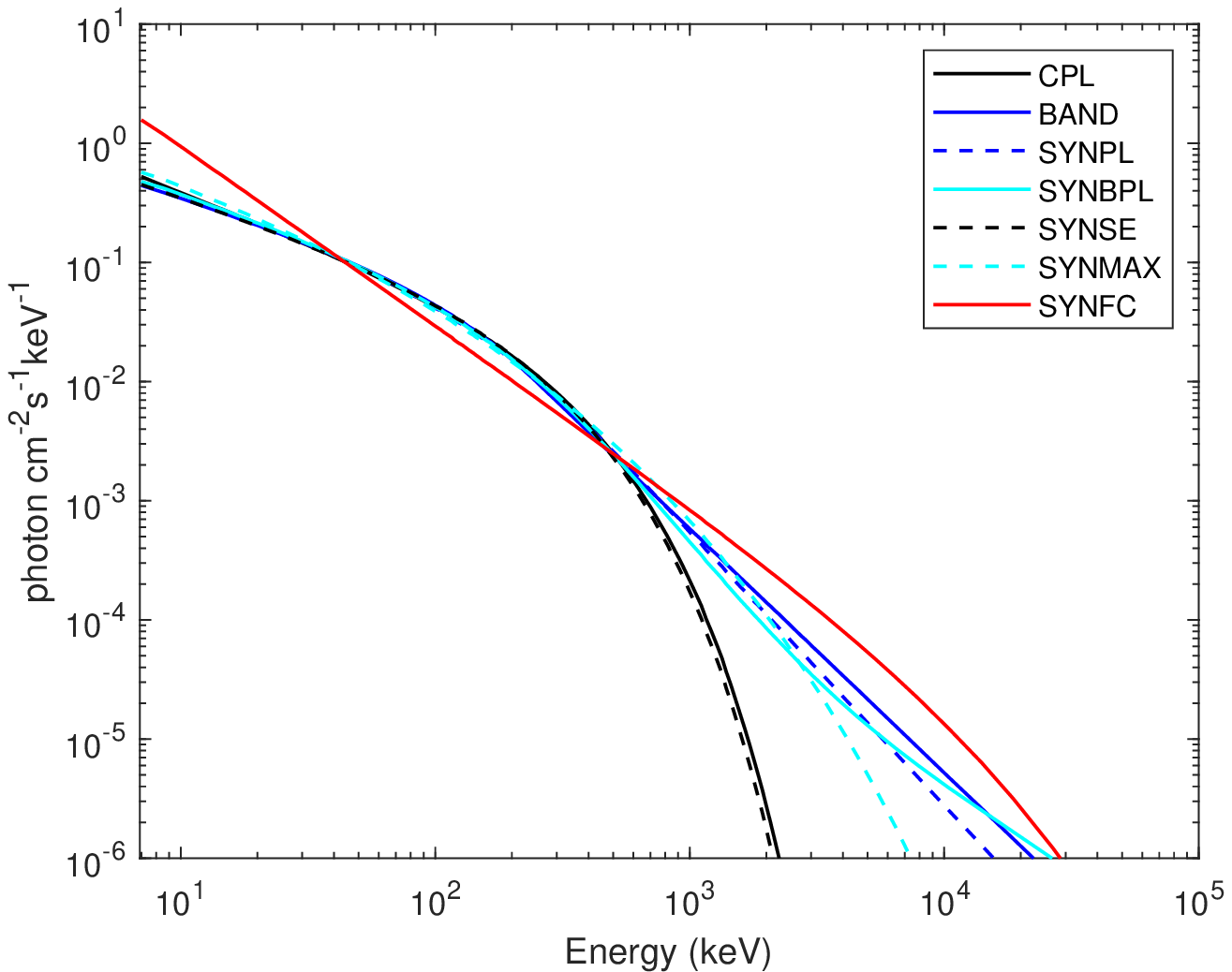}{0.45\textwidth}{(Photon spectra)}}

\caption{Comparison of observed count spectra with the fitted empirical models (up-left: CPL, up-middle: BAND) and synchrotron models (up-right: SYNPL; middle: SYNBPL, SYNMAX and SYNSE; bottom-left: SYNFC) for the time slice 8 of GRB 130821674. Bottom-right: comparison between the spectra of the seven models.}
\label{count-spectra}
\end{figure*}

\begin{figure*}
\centering
\gridline{\fig{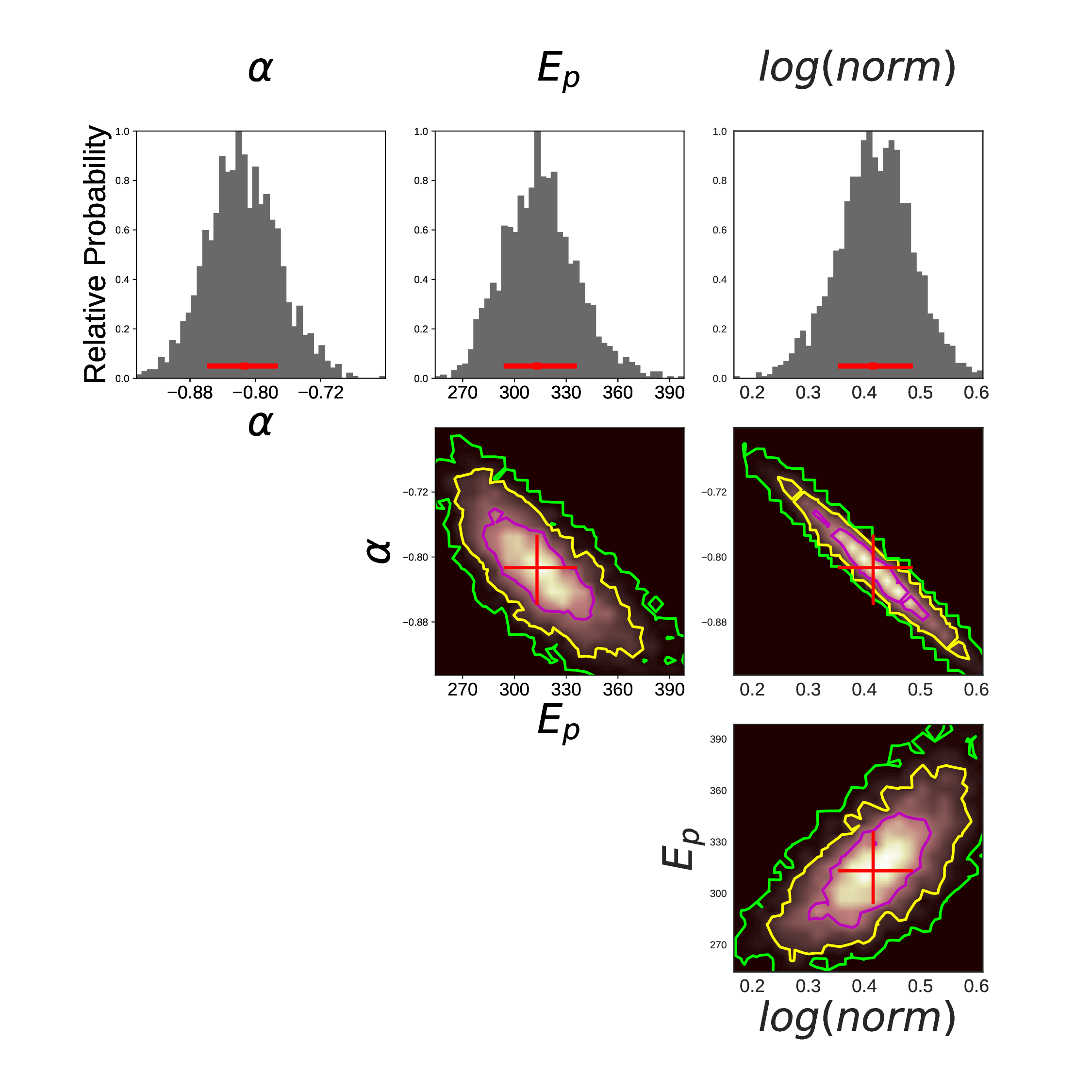}{0.35\textwidth}{(CPL)}
 \fig{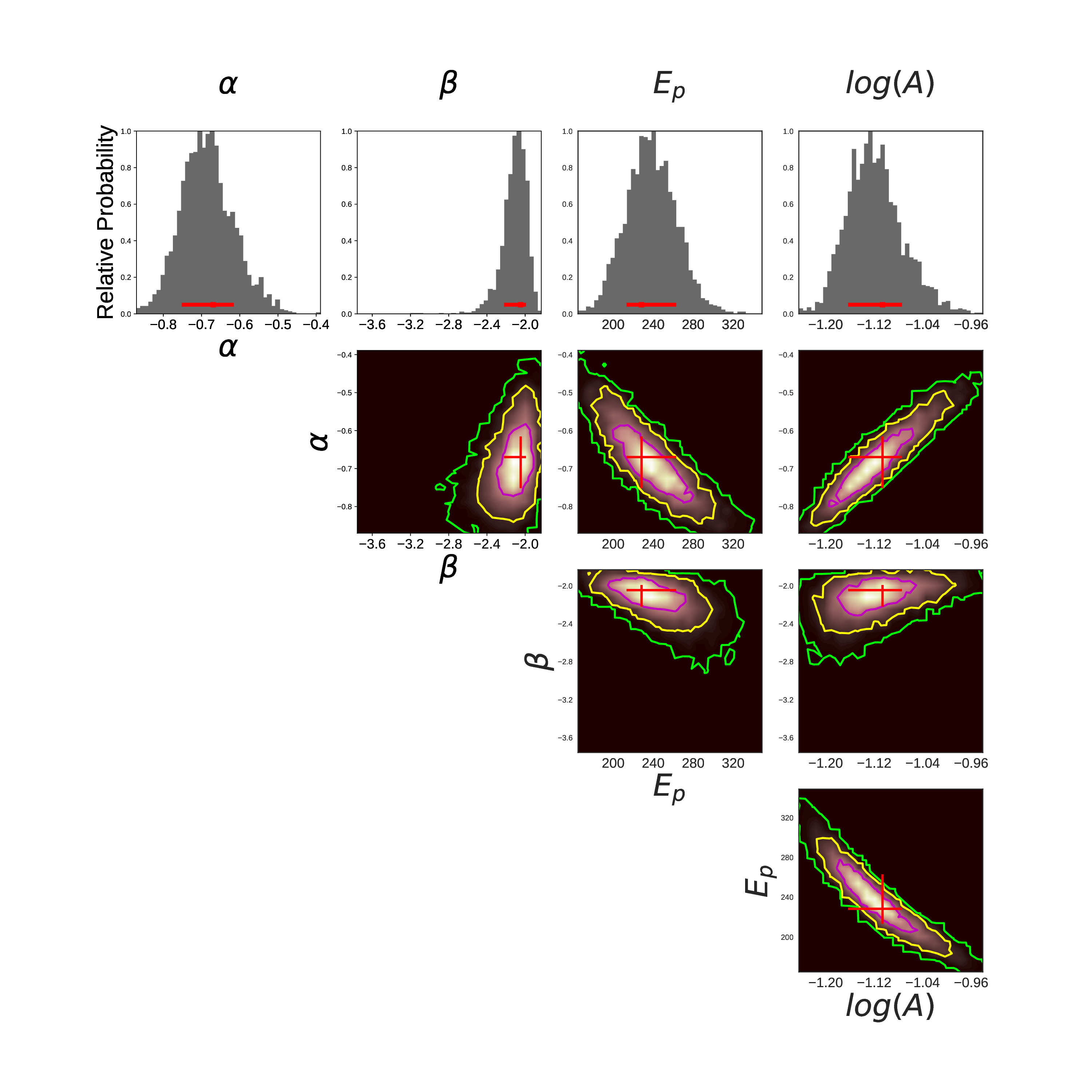}{0.35\textwidth}{(BAND)}
 \fig{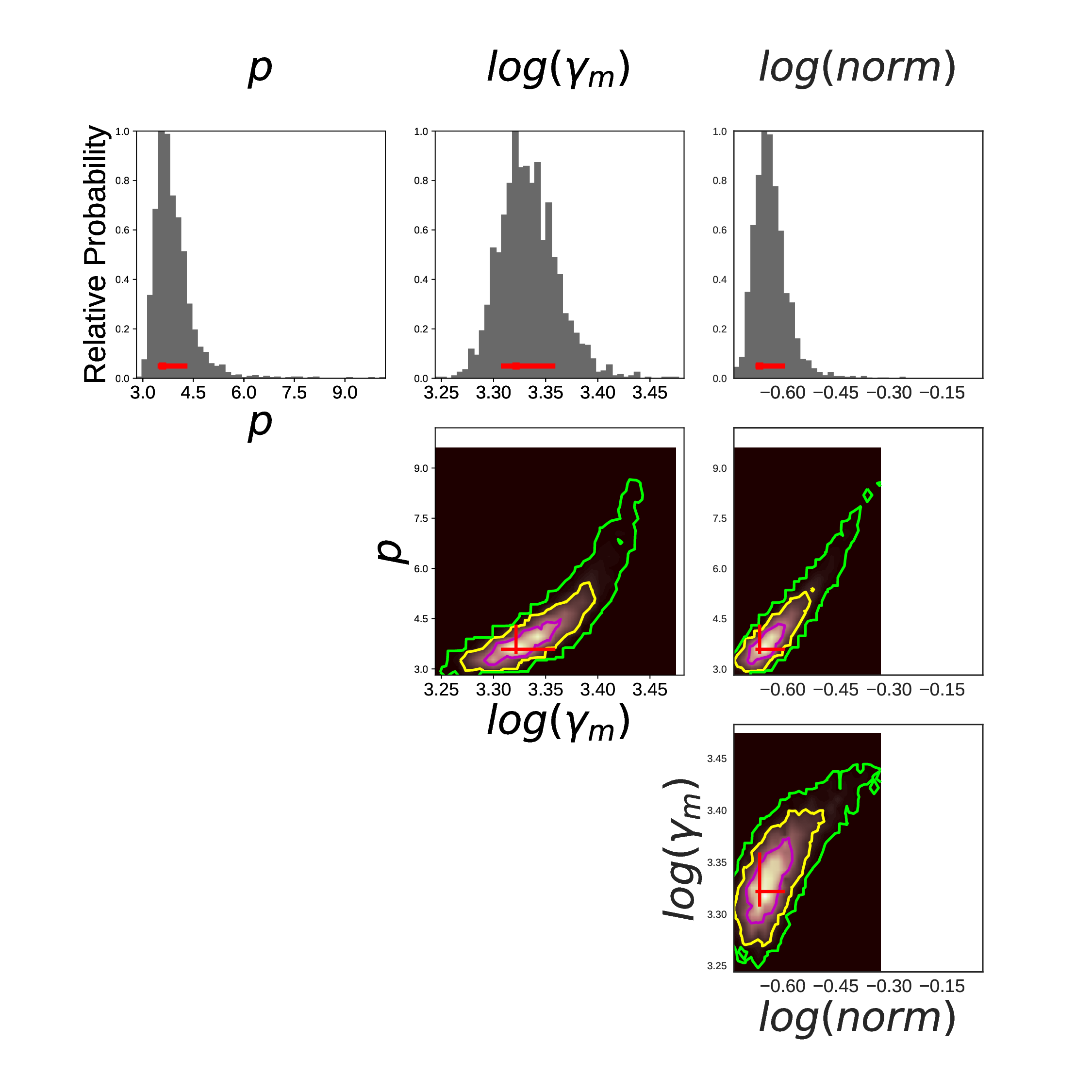}{0.35\textwidth}{(SYNPL)}}

\gridline{\fig{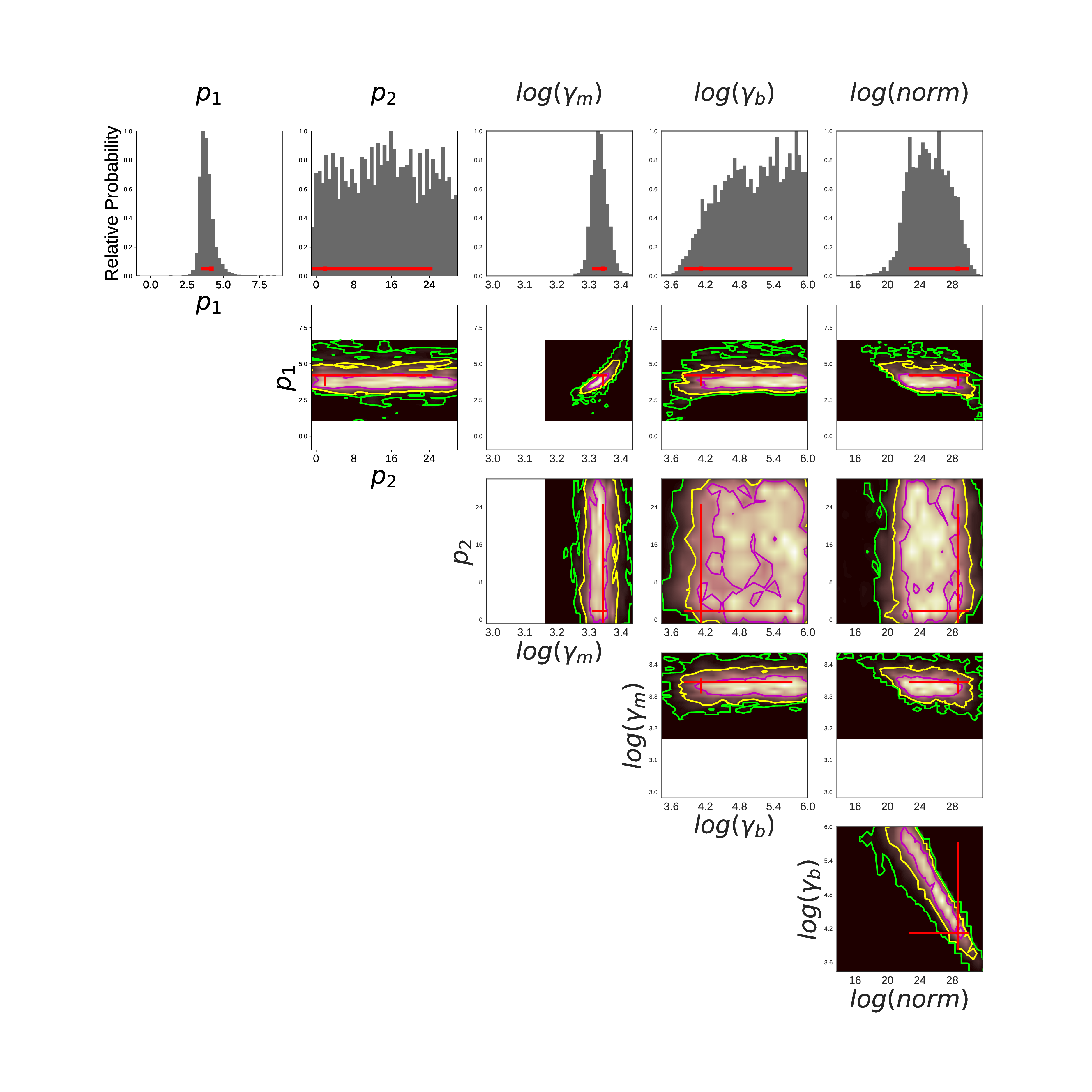}{0.35\textwidth}{(SYNBPL)}
 \fig{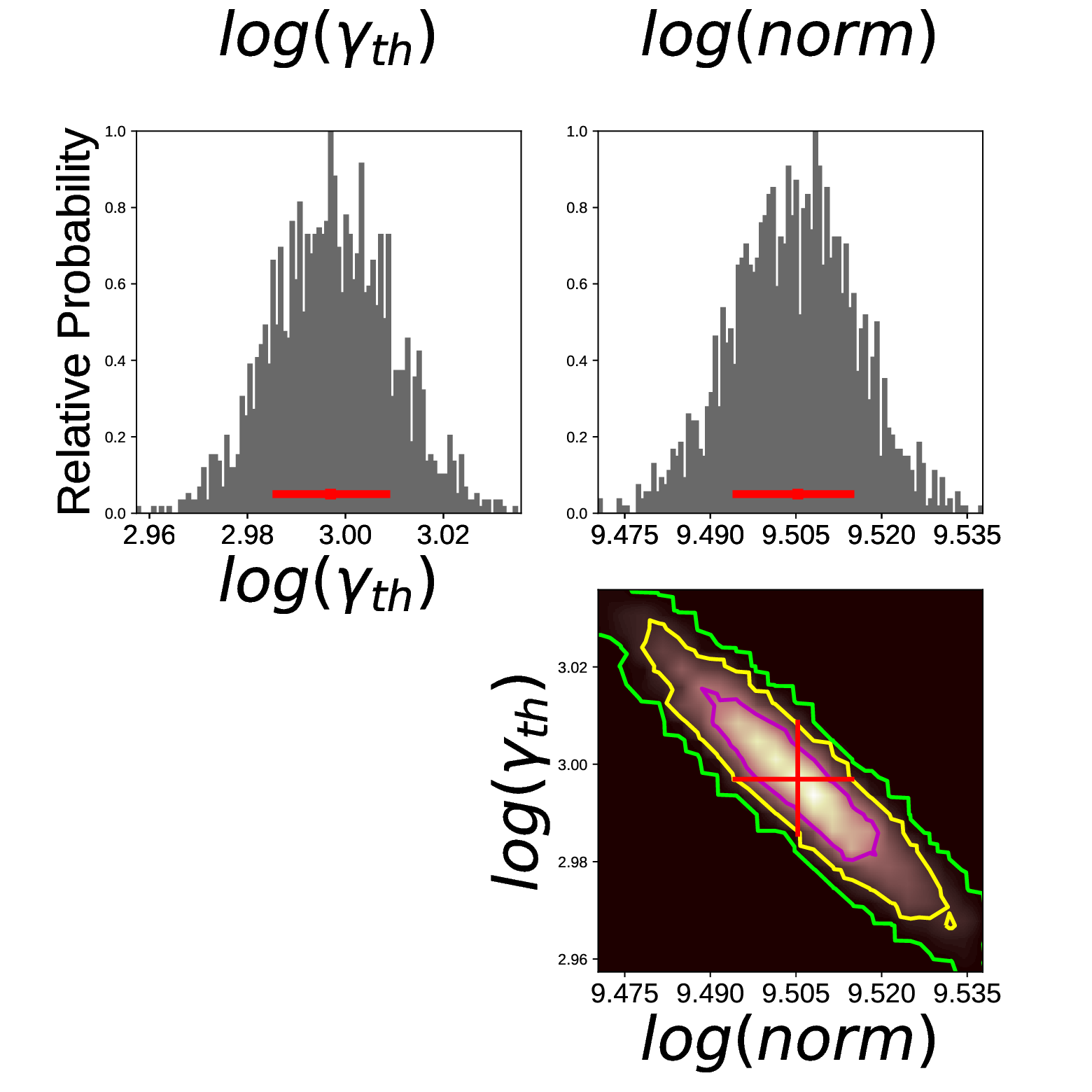}{0.35\textwidth}{(SYNMAX)}
 \fig{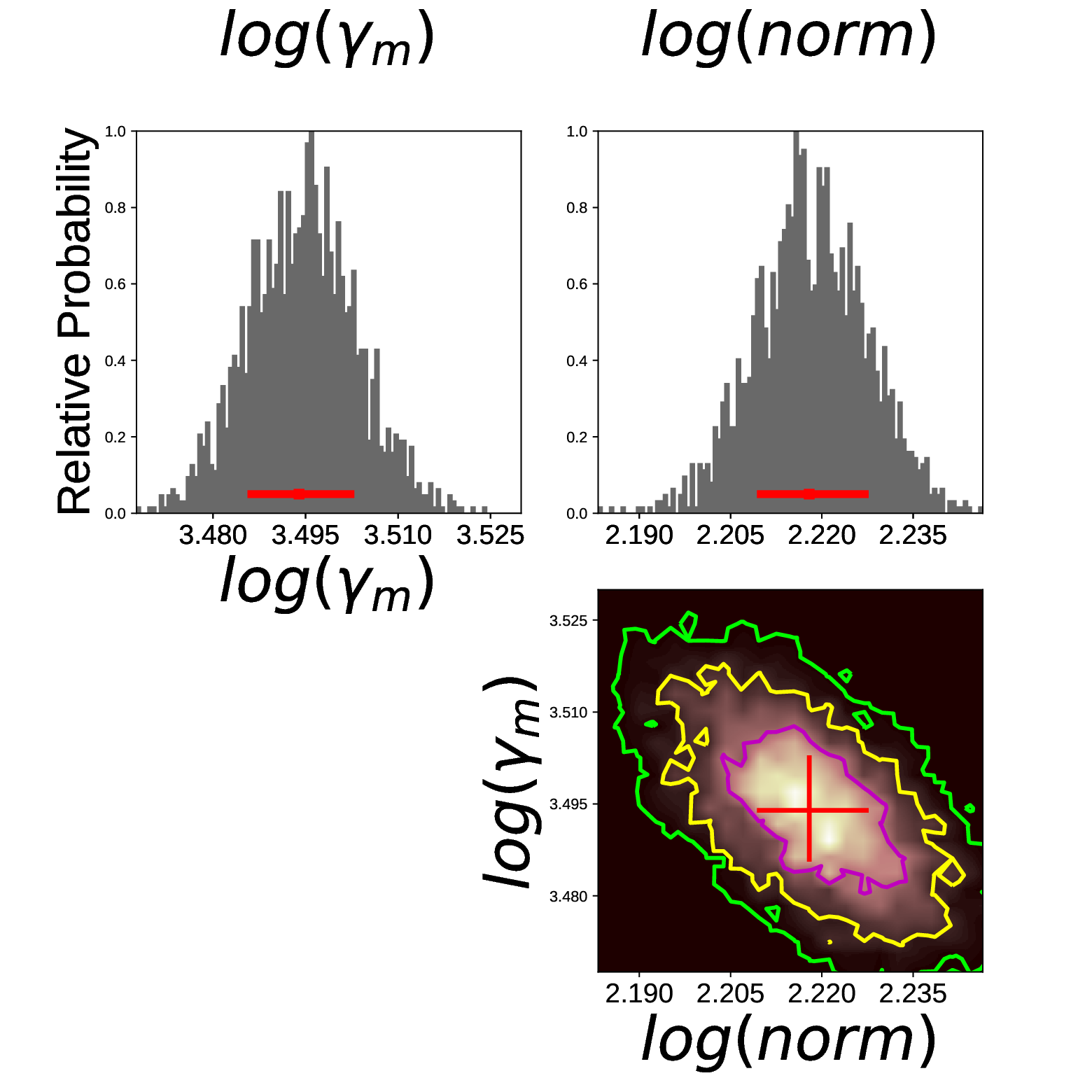}{0.35\textwidth}{(SYNSE)}}

\gridline{\fig{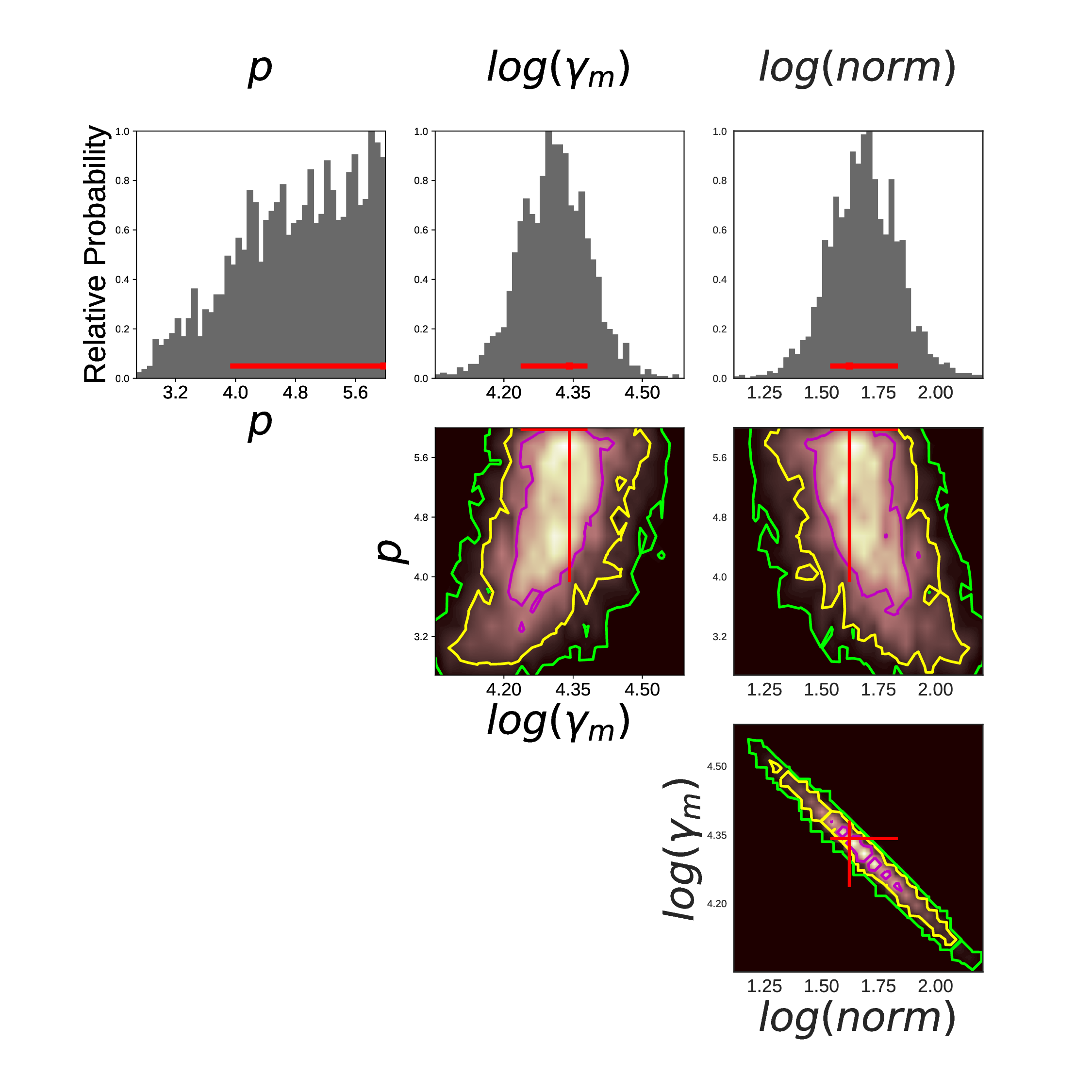}{0.35\textwidth}{(SYNFC)}}

 \caption{Parameter constraints of the empirical models (up-left: CPL, up-middle: BAND) and synchrotron models (up-right: SYNPL; middle: SYNBPL, SYNMAX and SYNSE; bottom: SYNFC) for the fit to the time slice 8 of GRB 130821674. Histograms and contours show the likelihood map of the parameter-constraint outputs by the McSpecFit package. Red crosses mark the best-fit values and 1 $\sigma$ error bars.}
 \label{contour}
\end{figure*}

An example of our spectral fit of seven models to one slice of data in GRB 130821674 is shown in Fig. \ref{count-spectra}, where one can see the fast cooling model fit presents a significant deviation in the low energies. The corresponding parameter constraints are shown in Fig. \ref{contour}. A whole-burst example that includes 15 slices, each fitted by seven models, is presented in Table 2.

\begin{figure*}
\centering
\includegraphics[width=20cm,height=8cm]{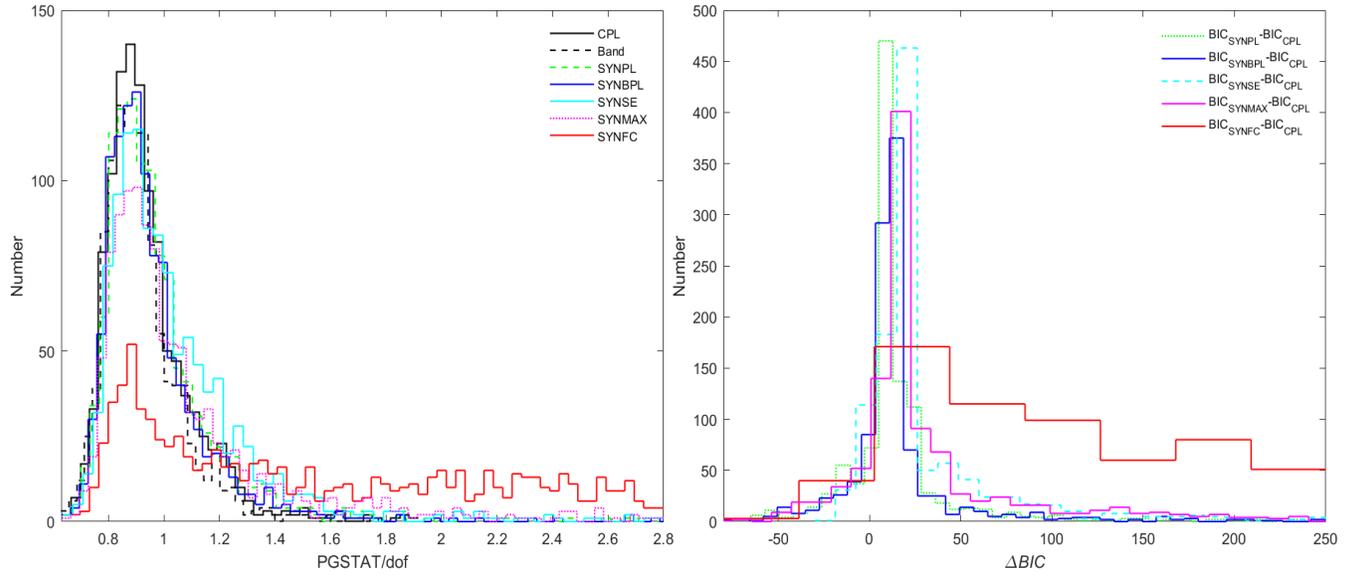}
 \caption{Left panel: distributions of the reduced goodness-of-fit (PGSTAT/dof) of all models. Right panel: distributions of the BIC difference between CPL and the physical models are shown.}
 \label{model_pgstat}
\end{figure*}

Figure \ref{model_pgstat} (left panel) sums up all the goodness of the fit to all different models. We found that the CPL model can successfully fit the 91\% of the observed spectra with goodness-of-fit within range of [0.75,1.25], and the reduced goodness-of-fit (PGSTAT/d.o.f) distribution is much less spreading than all other models. We thus treat CPL as the representative empirical model to best describe the observed data. Other physical models are further validated by comparing the CPL model using Bayesian Information Criteria (BIC). For each of the 1117 spectra sets, we calculate the BIC difference between each physical model fit and the CPL model fit. Finally, we plot the distributions of the difference in Fig. \ref{model_pgstat} (right panel). From a statistical point of view, our results show that the SYNPL model, with the minimal $\Delta$BIC compared to other physical models, is the best physical model approaching the empirical CPL model. The SYNBPL, SYNMAX and SYNSE models are systemically unfavorable compared to the SYNPL model, yet somehow indistinguishable from their $\Delta$BIC values. Moreover, our results suggest that the classical SYNFC model is incapable of resembling the observed data compared to the CPL model.

The median values and standard deviations of the distribution of the model parameters are summarized in Table 3. The distribution of the low energy index, $\alpha$, which is in the empirical model (such as CPL and BAND) fits, peaks at slightly less than $-2/3$, which is the synchrotron slope below the minimum injected energy without cooling. The peak energy, $E_{p}$, of the BAND model, is lower than that of the CPL model (see Table 3). This result is consistent with the statistical results of previous GBM spectral catalogs (e.g., Yu et al. 2016, 2019). The high-energy index $\beta$ of the BAND spectrum is smaller than the previous spectral catalogs. A small part of the sample's $\beta$ is very small ($-$10 to $-$4), indicating that this part of the energy spectrum is a high-energy cut-off single power-law spectrum.

\begin{figure*}
\centering
	\includegraphics[width=5.8cm,height=4.5cm]{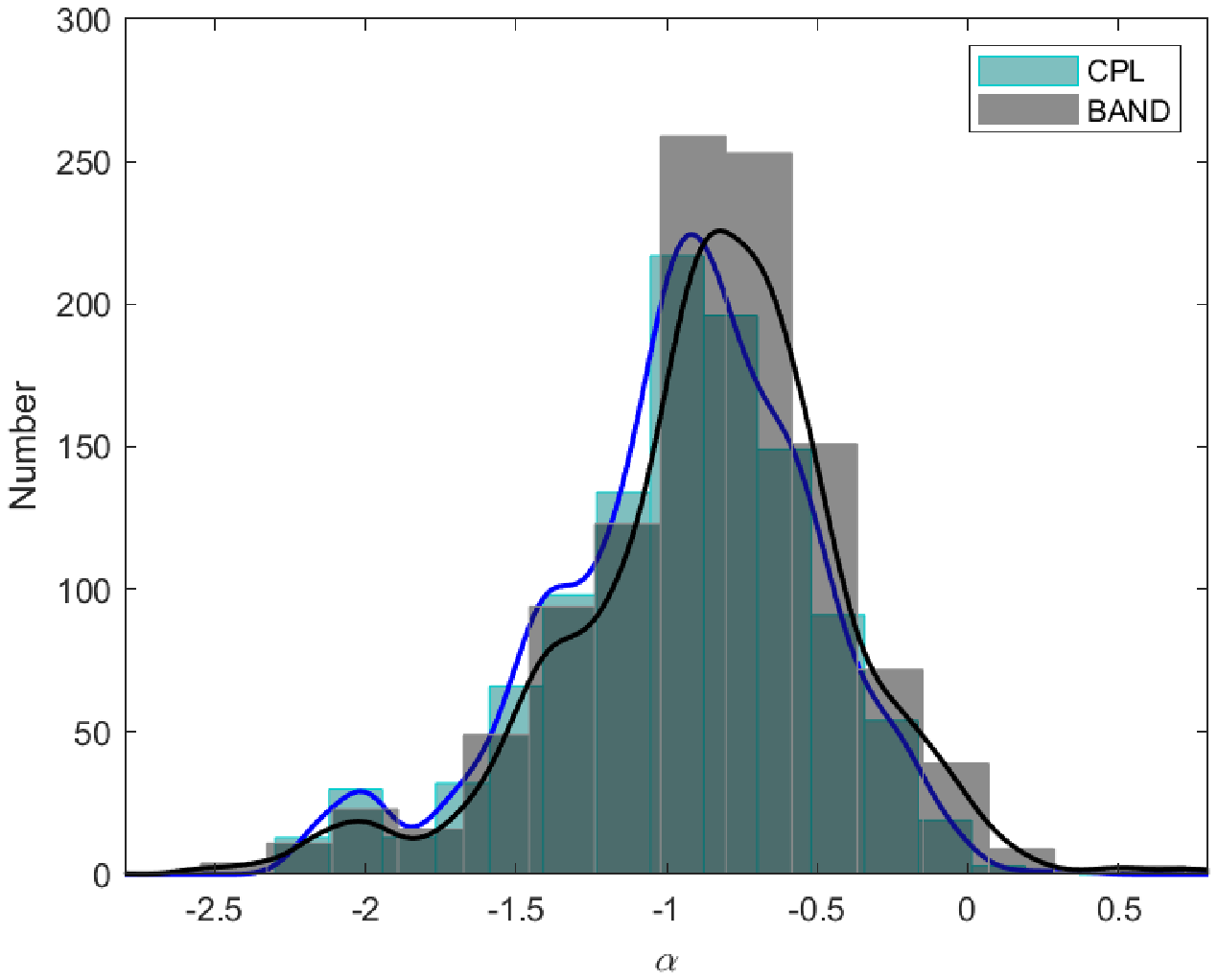}
 \includegraphics[width=5.8cm,height=4.5cm]{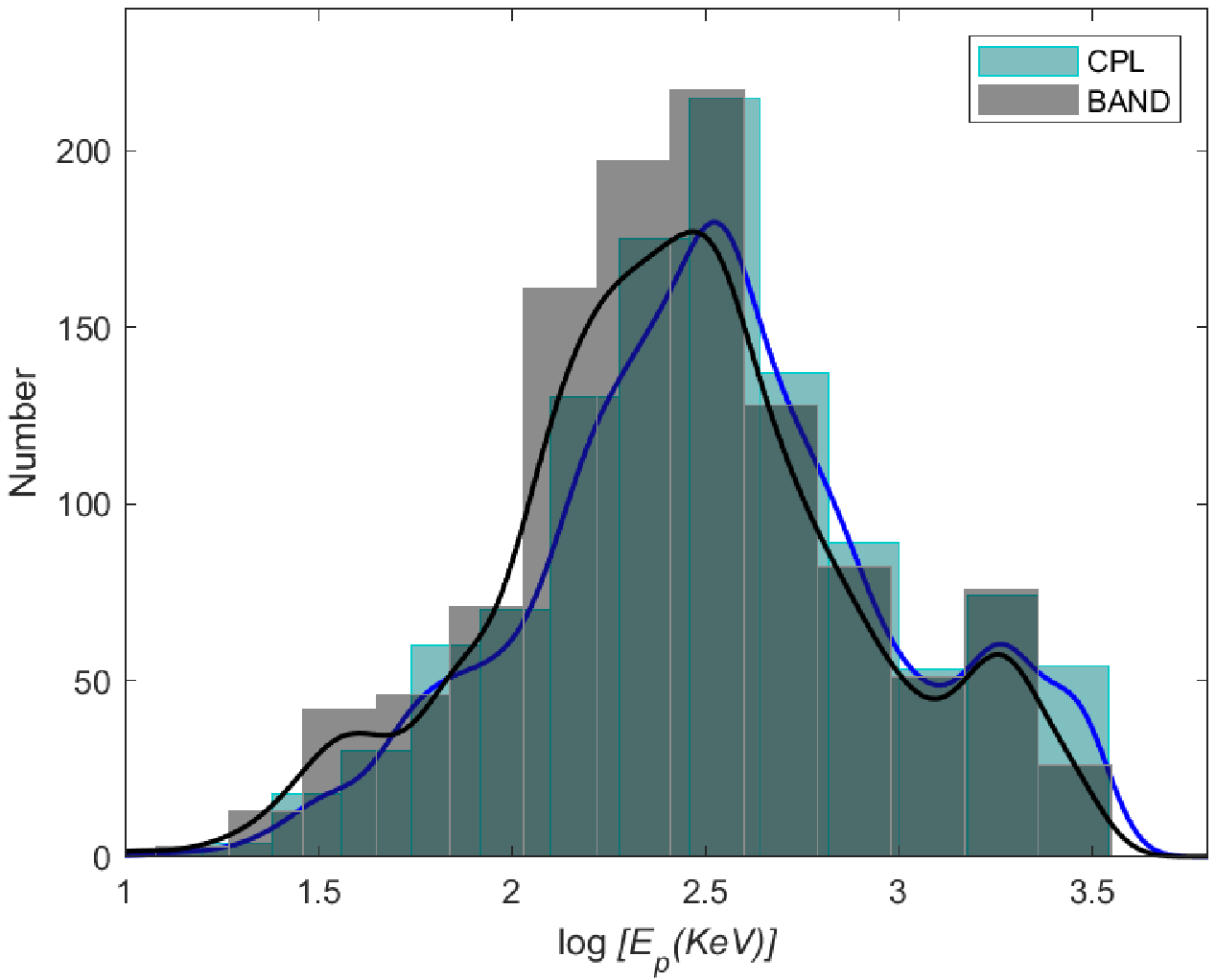}
 \includegraphics[width=5.8cm,height=4.5cm]{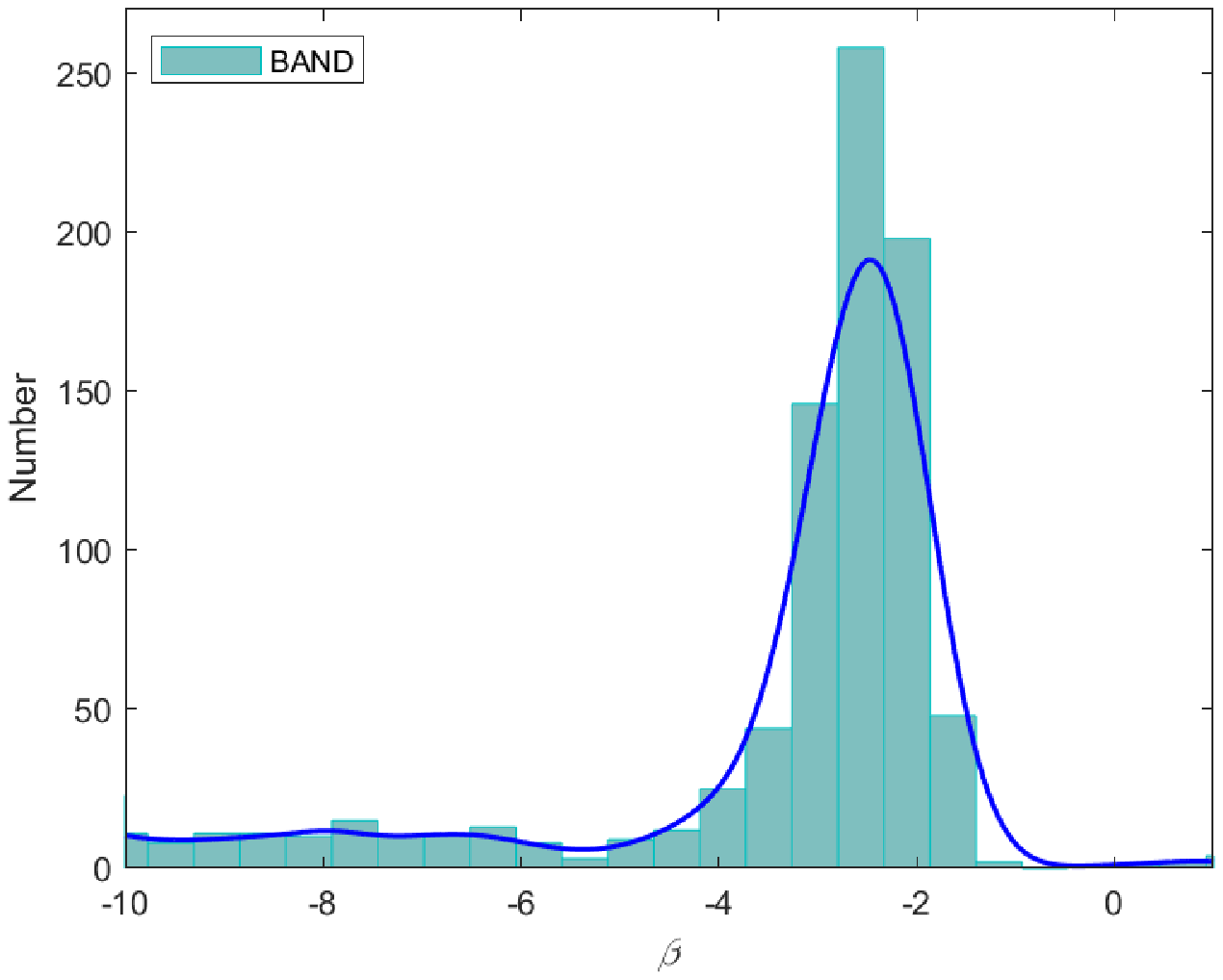}
 \includegraphics[width=5.8cm,height=4.5cm]{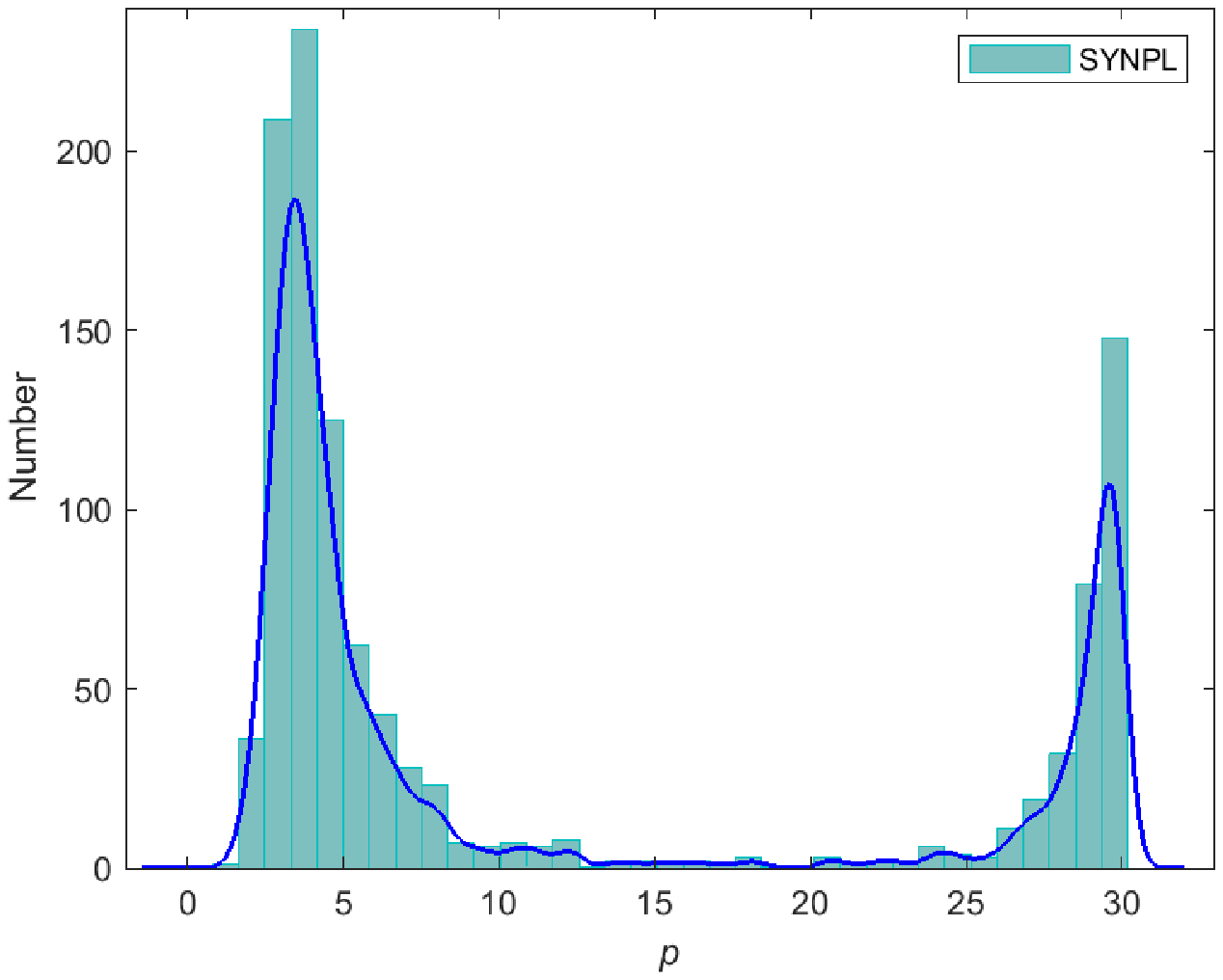}
 \includegraphics[width=5.8cm,height=4.5cm]{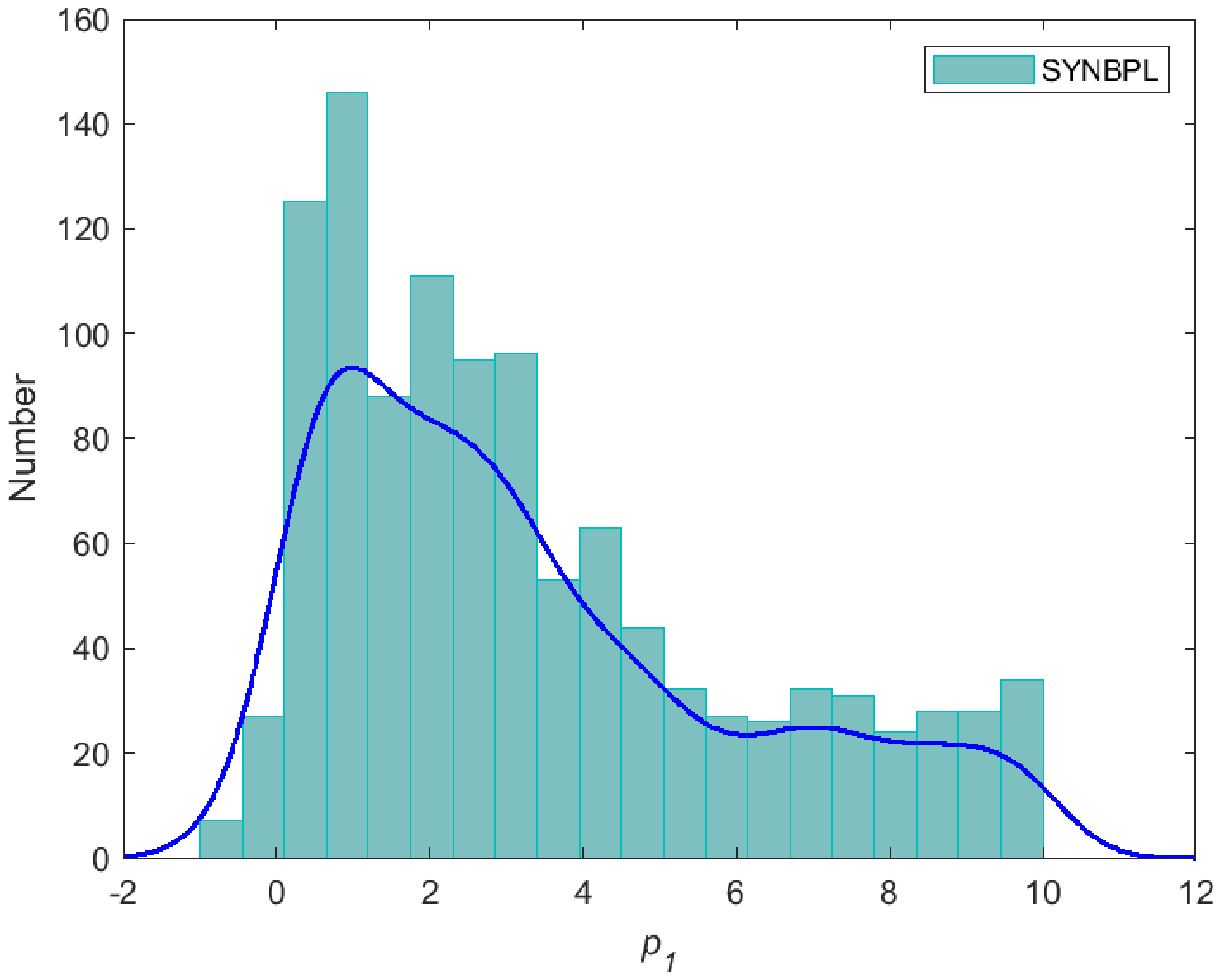}
 \includegraphics[width=5.8cm,height=4.5cm]{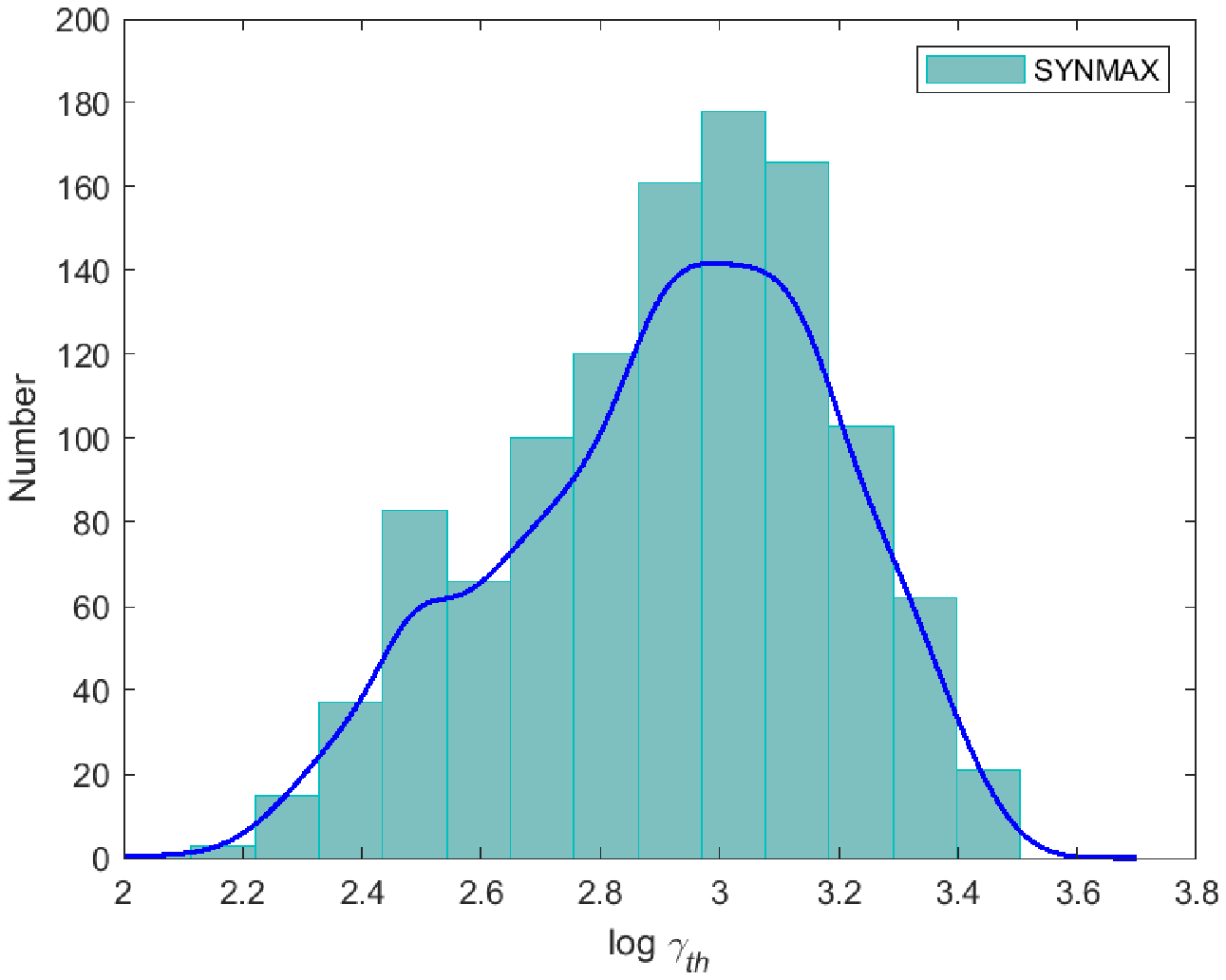}
 \includegraphics[width=6cm,height=4.5cm]{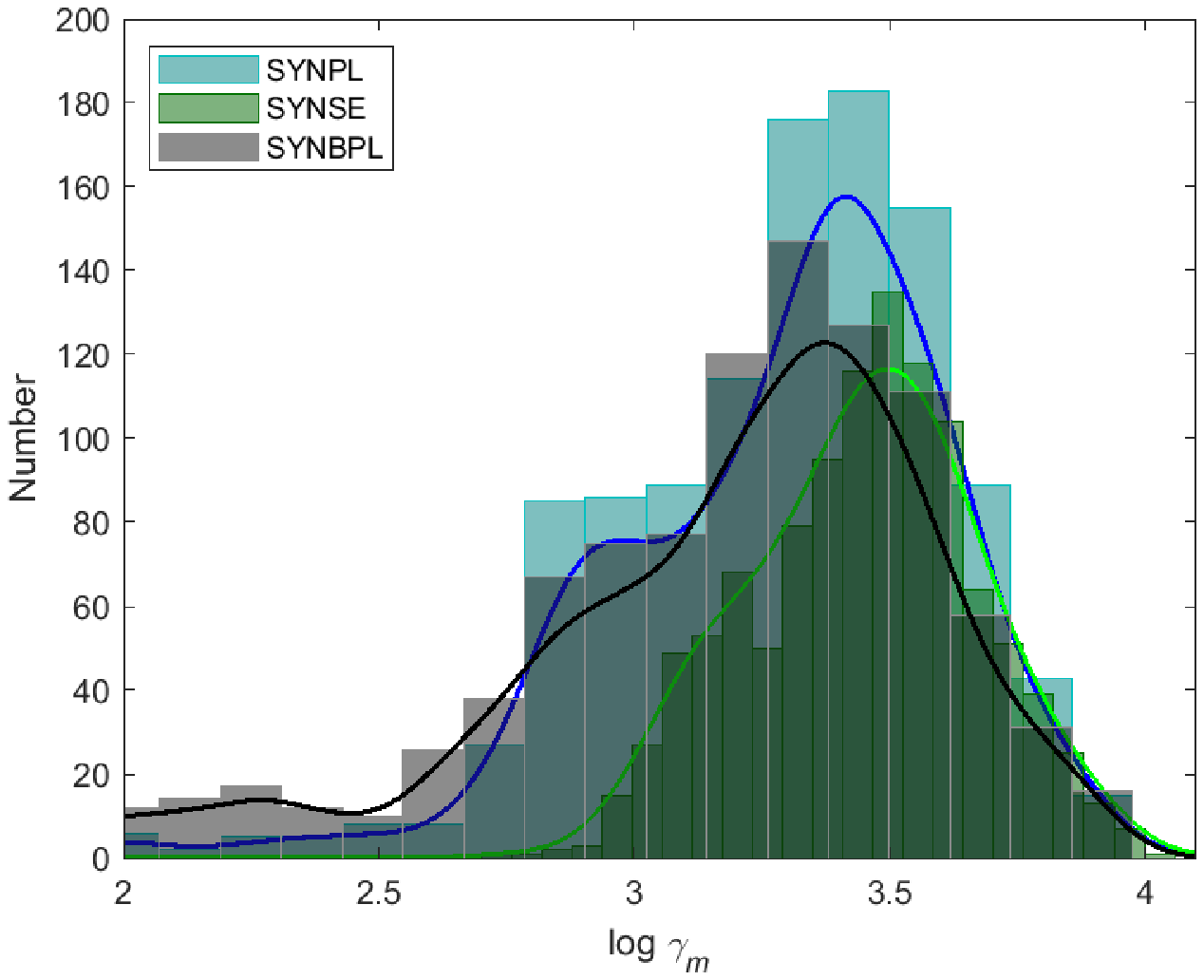}
 \includegraphics[width=6cm,height=4.5cm]{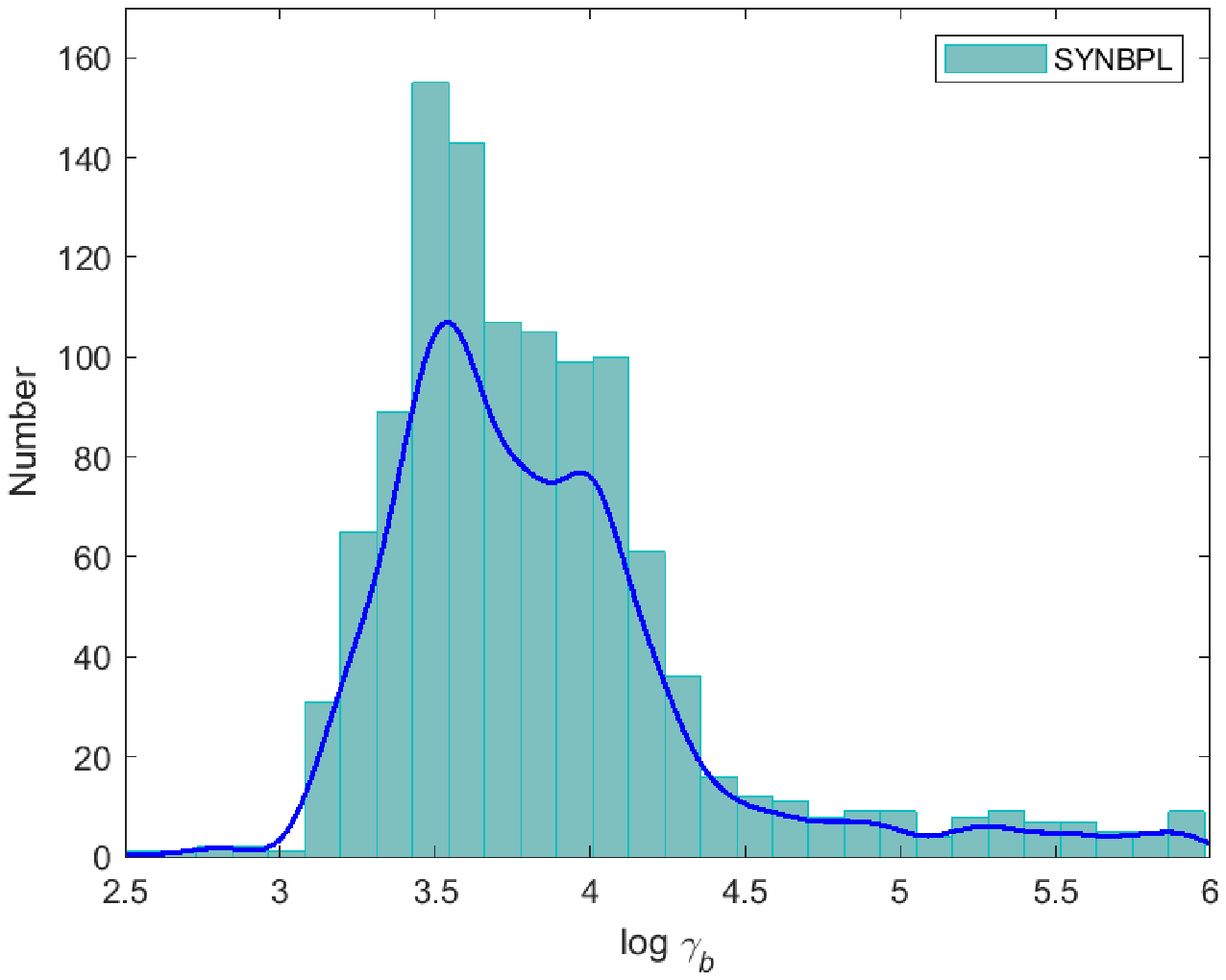}
 \caption{Distributions of each model parameter are shown, including $\alpha$ and $log (E_{p})$ of CPL and BAND; $\beta$ of BAND; $p$ and $log(\gamma_{m})$ of SYNPL; $p_{1}$, $log(\gamma_{m})$ and $log(\gamma_{b})$ of SYNBPL; $log(\gamma_{m})$ of SYNSE; $log(\gamma_{th})$ of SYNMAX. The SYNFC is not included because it fits the data poorly. The curves represent the kernel density estimation (KDE) of the distributions, where Gaussian kernels are used.}
 \label{para-dis}
\end{figure*}

The distribution of the spectral index $p$ of the SYNPL model has two peaks (see Fig. \ref{para-dis}). About 3/4 of the spectral sample peaks at $\sim 3.6$, while the left 1/4 peaks at $\sim 29.6$, which suggests these spectra actually can be fitted by the SYNSE model. The distribution of $p_1$ of the SYNBPL model is very wide, varying from $\sim 0$ to $\sim 10$. This may be attributed to the poor constraint on the parameters of the SYNBPL model. The distributions of the minimum injection energy ($\gamma_{m}$) for the SYNSE, SYNPL, and SYNBPL models peak at $\sim 10^{3.4}$, which corresponds to the synchrotron peak of $\sim 100$ keV, while the peak of $\gamma_{b}$ for the SYNBPL model is $10^{3.8}$ corresponding to the synchrotron peak of $\sim 400$ keV (see Table 3).

\begin{figure*}
\centering
 \includegraphics[width=8cm,height=6cm]{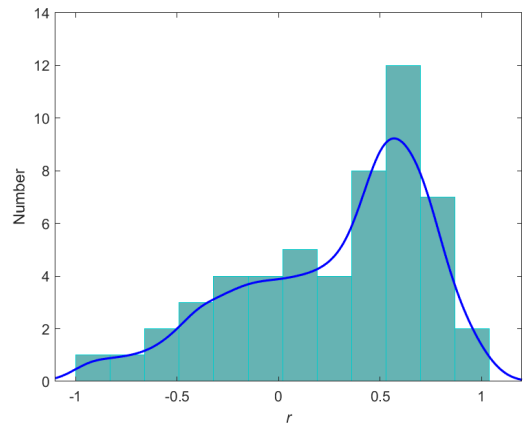}
 \caption{Distribution of the correlation coefficient ($r$) between the electron distribution index of the SYNPL model and the flux of 53 GRBs.}
 \label{r-dis}
\end{figure*}

The electron power-law index, $p$, is a critical parameter in the SYNPL model. Interestingly, a correlation between the flux $F$ and the index $p$ is found in our analysis. Using Spearman’s rank correlation analysis method, we calculate the linear correlation coefficients ($r$) and chance probabilities ($P$) for each burst and evaluate the correlation of $F-p$. The distribution of $r$ is presented in Fig. \ref{r-dis}. One can find that the distribution of $r$ is somewhat diverse, but clearly, there is a peak around $r\sim0.6$. It is $r>0.4$ for more than a half bursts in our samples, suggesting strong positive correlations between $F$ and $p$. We list 6 cases in Fig. \ref{p-evol}. The evolution of the electron power-law index traces the light curve for these bursts.

\section{Summary and Implications}

In this paper, using several synchrotron models with different electron distributions and two empirical models, a time-resolved spectroscopy analysis is performed for 53 long bright bursts of \textit{Fermi}/GBM, including a total of 1117 time-resolved spectra. According to the Bayesian information standard, we compare the fitting results of each model. We find that: (1) most of the GRB energy spectra are well fitted with the synchrotron models, except the SYNFC model, and the SYNPL model is better than the SYNSE, SYNMAX, and SYNBPL models based on the BIC criterion; (2) a strong correlation between $F-p$ is found for the SYNPL model, i.e., the electron power-law index in more than a half bursts traces their flux; (3) the empirical models are comparably good to synchrotron models when fitting to the observed spectra.

Our results support the previous conclusion that the synchrotron mechanism can be responsible for the observed GRB spectra (e.g., Oganesyan et al. 2017; Beniamini et al. 2018; Ravasio et al. 2018 and 2019; Burgess et al. 2014 and 2020). However, the traditional fast cooling synchrotron model is not favored in our fits. Our findings that SYNPL model can fit the data suggest that the heated electrons in the emission region is not significantly cooled or $\gamma_c$ is right around $\gamma_m$. These suggest that GRBs may result from magnetic dissipation at much larger radii with a weak magnetic field instead of the internal shock at smaller radii with a large magnetic field (e.g., Kumar et al. 2007; Zhang \& Yan 2011).

We note that our model comparison is subject to the caveat that the forward-folding method, which requires the input models to be first convolved with the instrumental response (RSP) matrix before compared with the observed count spectrum, may be incapable of discriminating between various models due to the nonlinear nature of the RSP matrix.

\begin{figure*}
 \includegraphics[width=5.8cm,height=4.2cm]{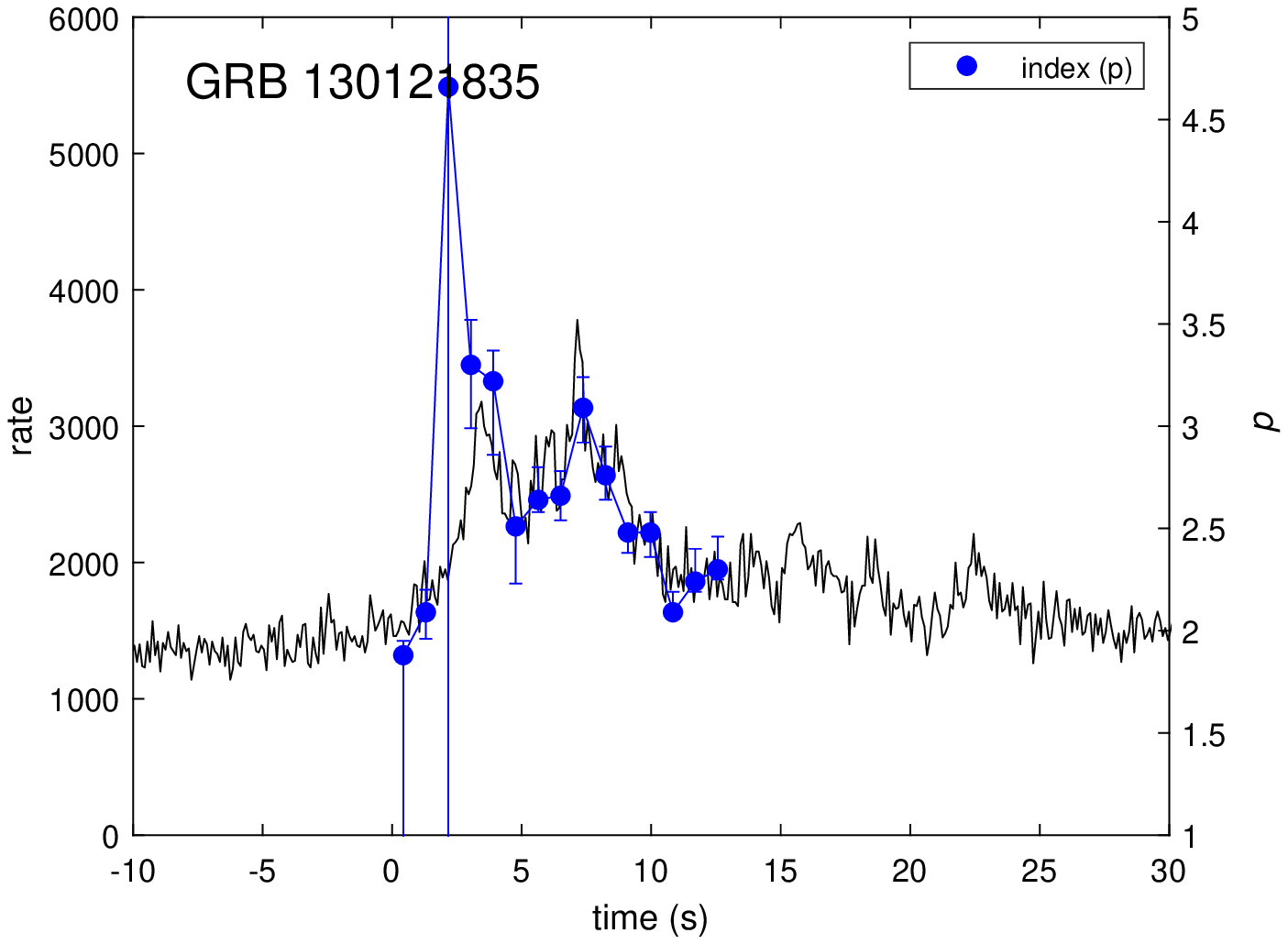}
 \includegraphics[width=5.8cm,height=4.2cm]{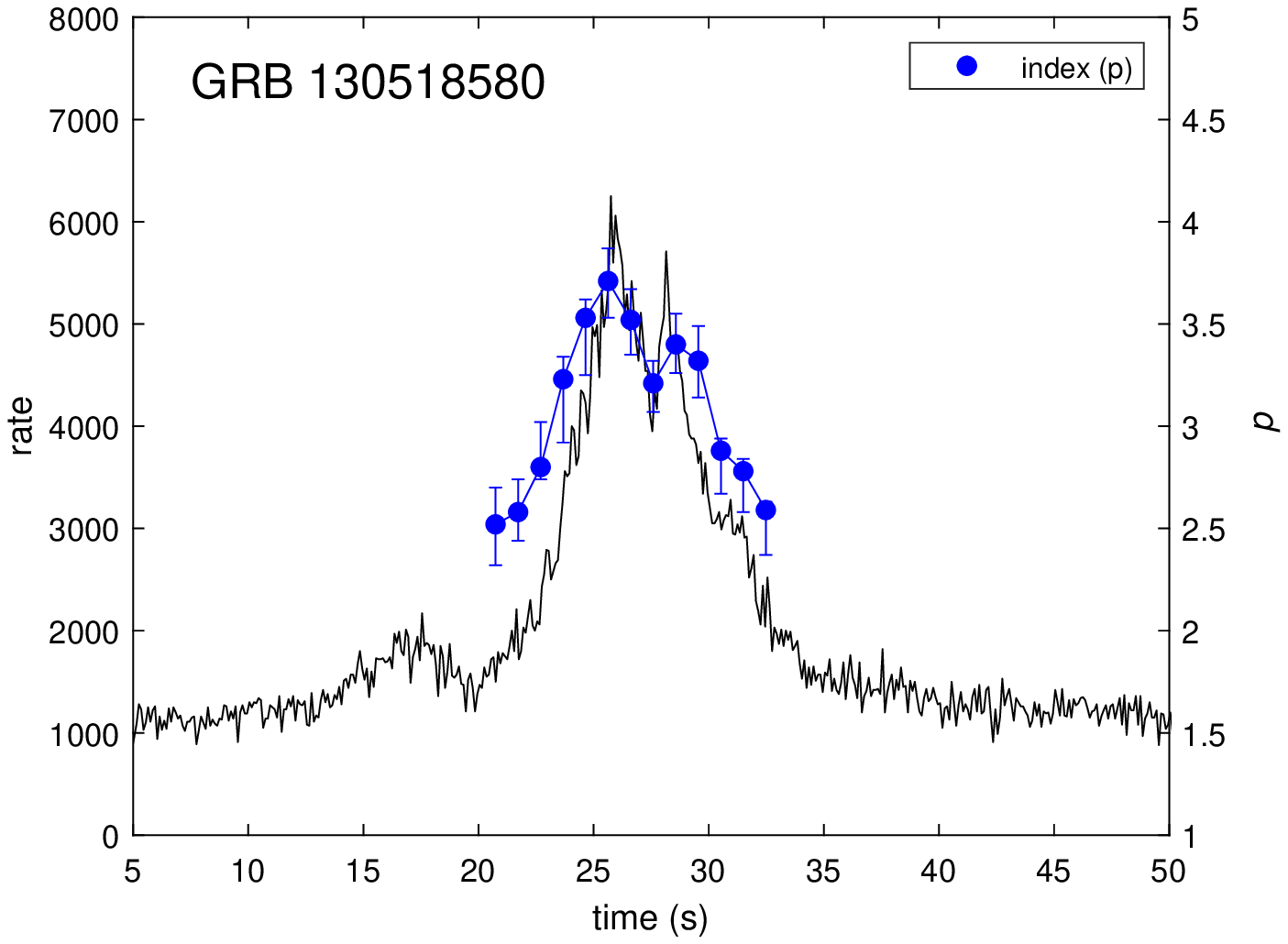}
 \includegraphics[width=5.8cm,height=4.2cm]{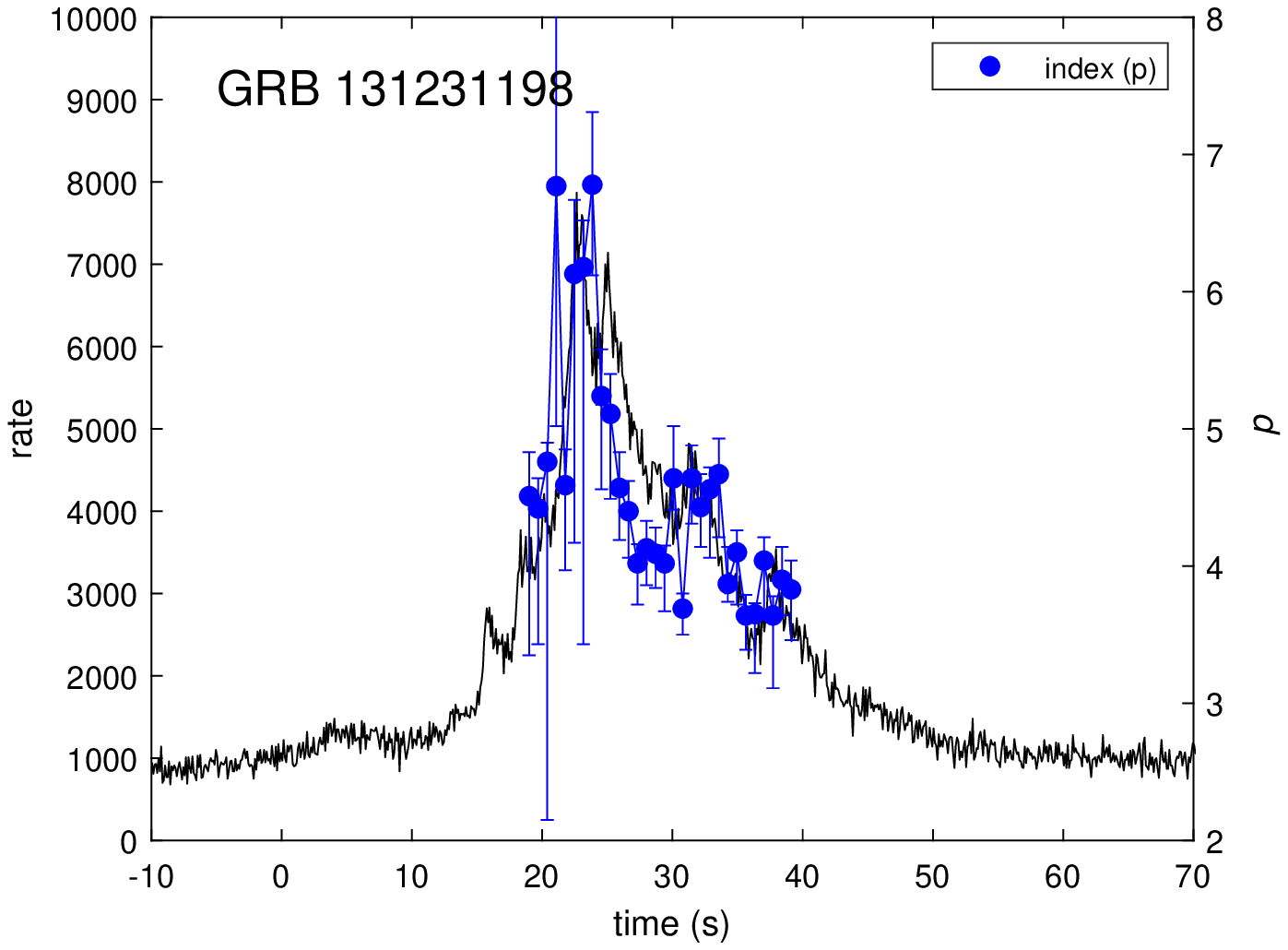}

 \includegraphics[width=5.8cm,height=4.2cm]{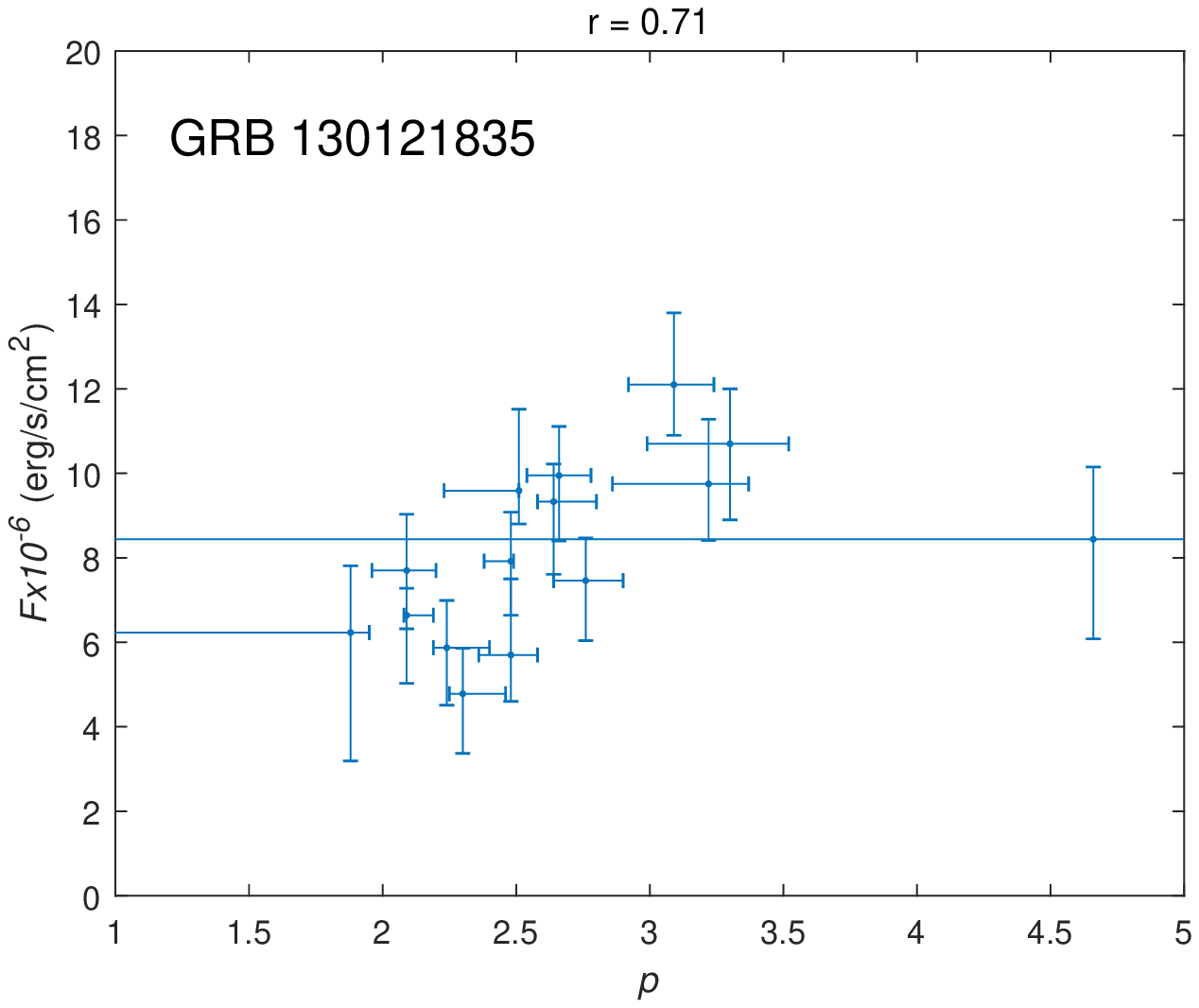}
 \includegraphics[width=5.8cm,height=4.2cm]{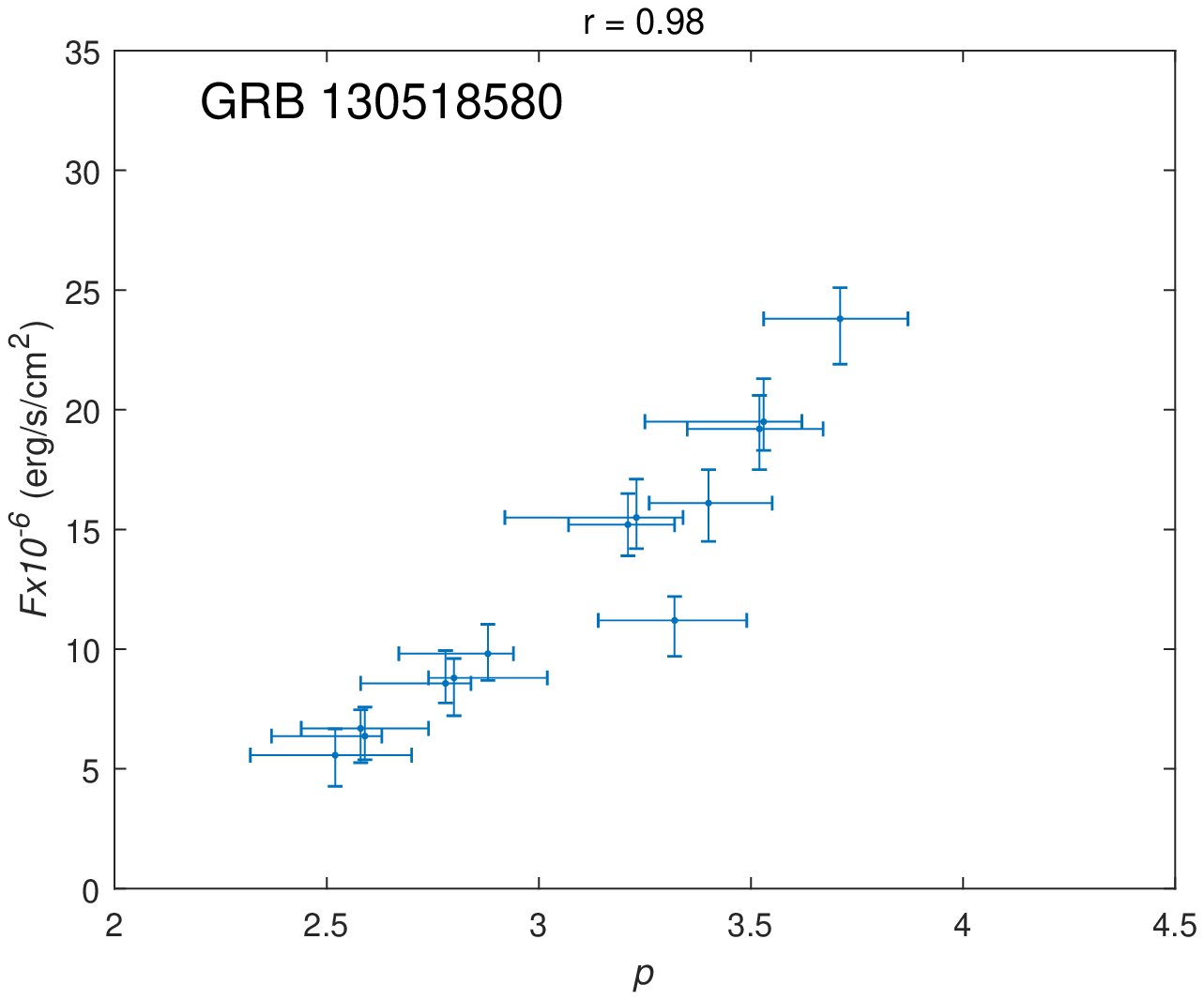}
 \includegraphics[width=5.8cm,height=4.2cm]{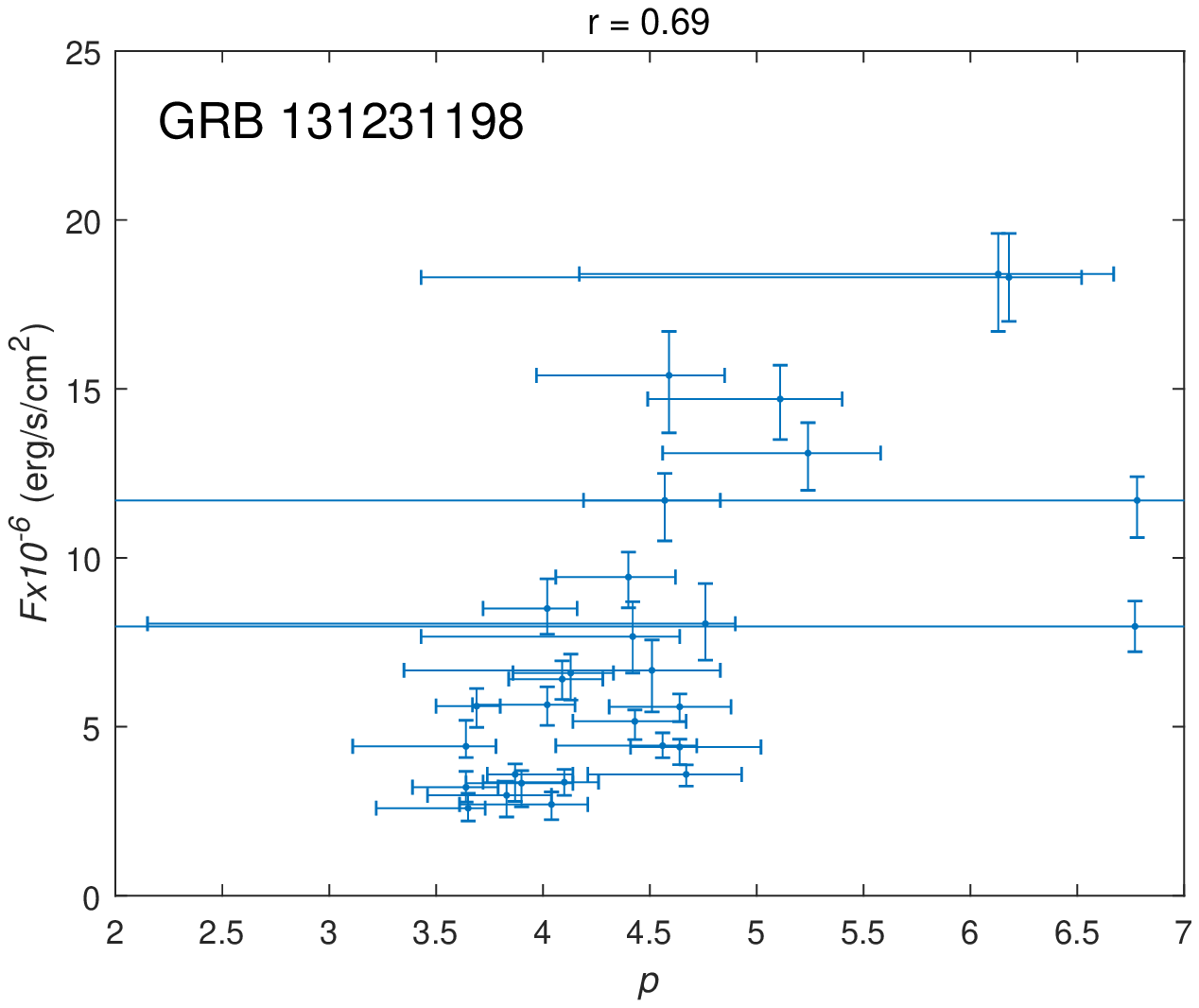}

 \includegraphics[width=5.8cm,height=4.2cm]{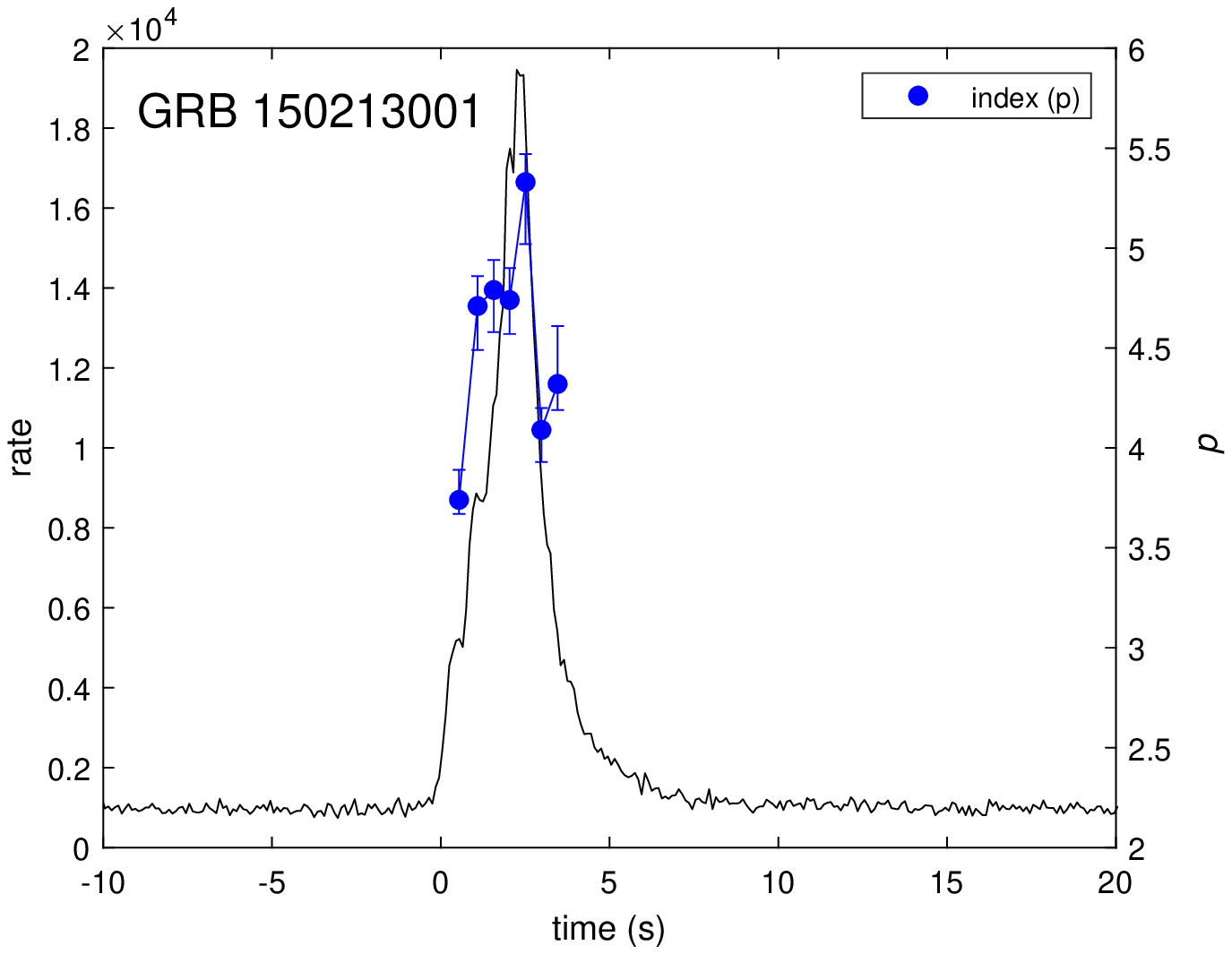}
 \includegraphics[width=5.8cm,height=4.2cm]{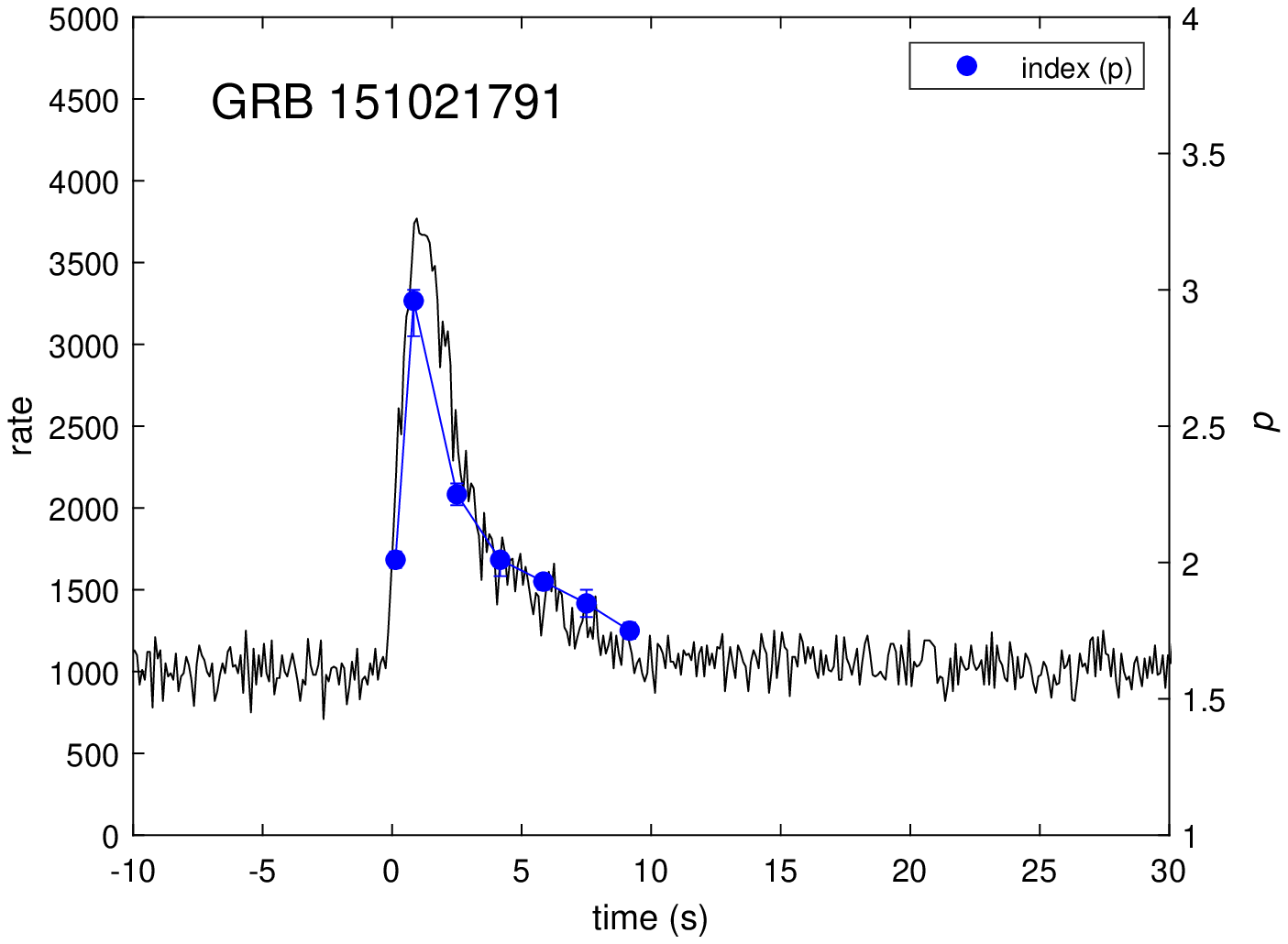}
 \includegraphics[width=5.8cm,height=4.2cm]{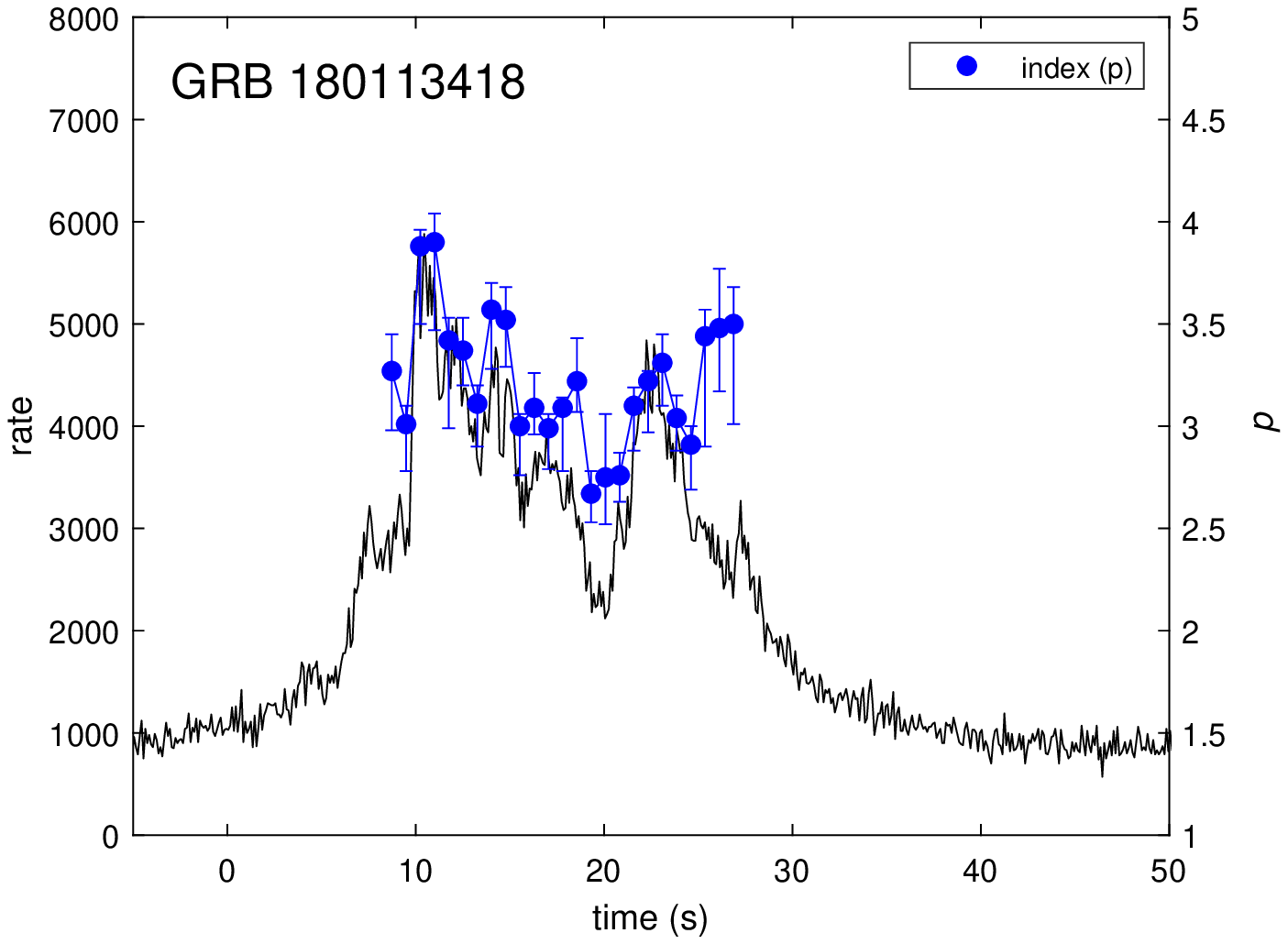}

 \includegraphics[width=5.8cm,height=4.2cm]{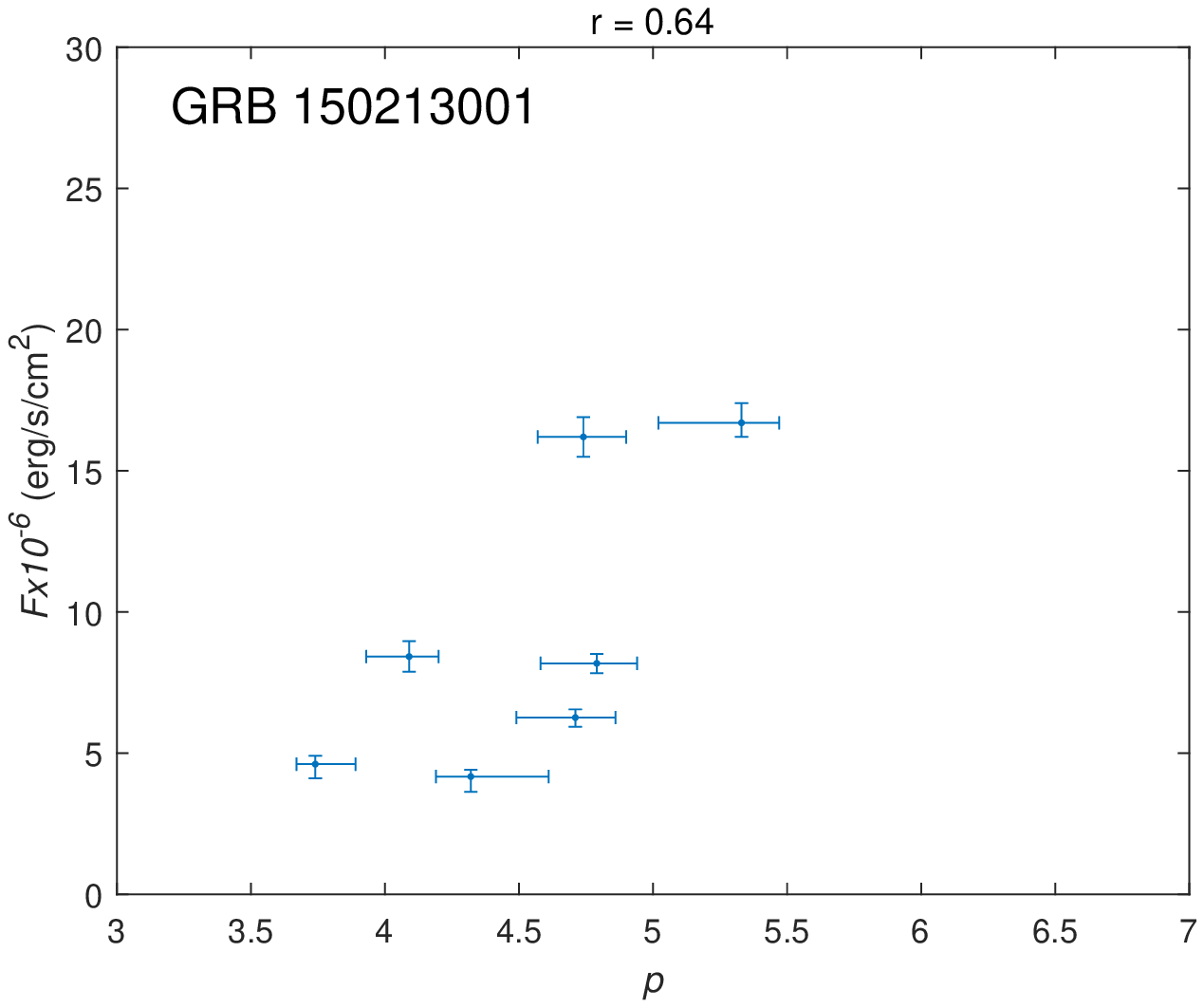}
 \includegraphics[width=5.8cm,height=4.2cm]{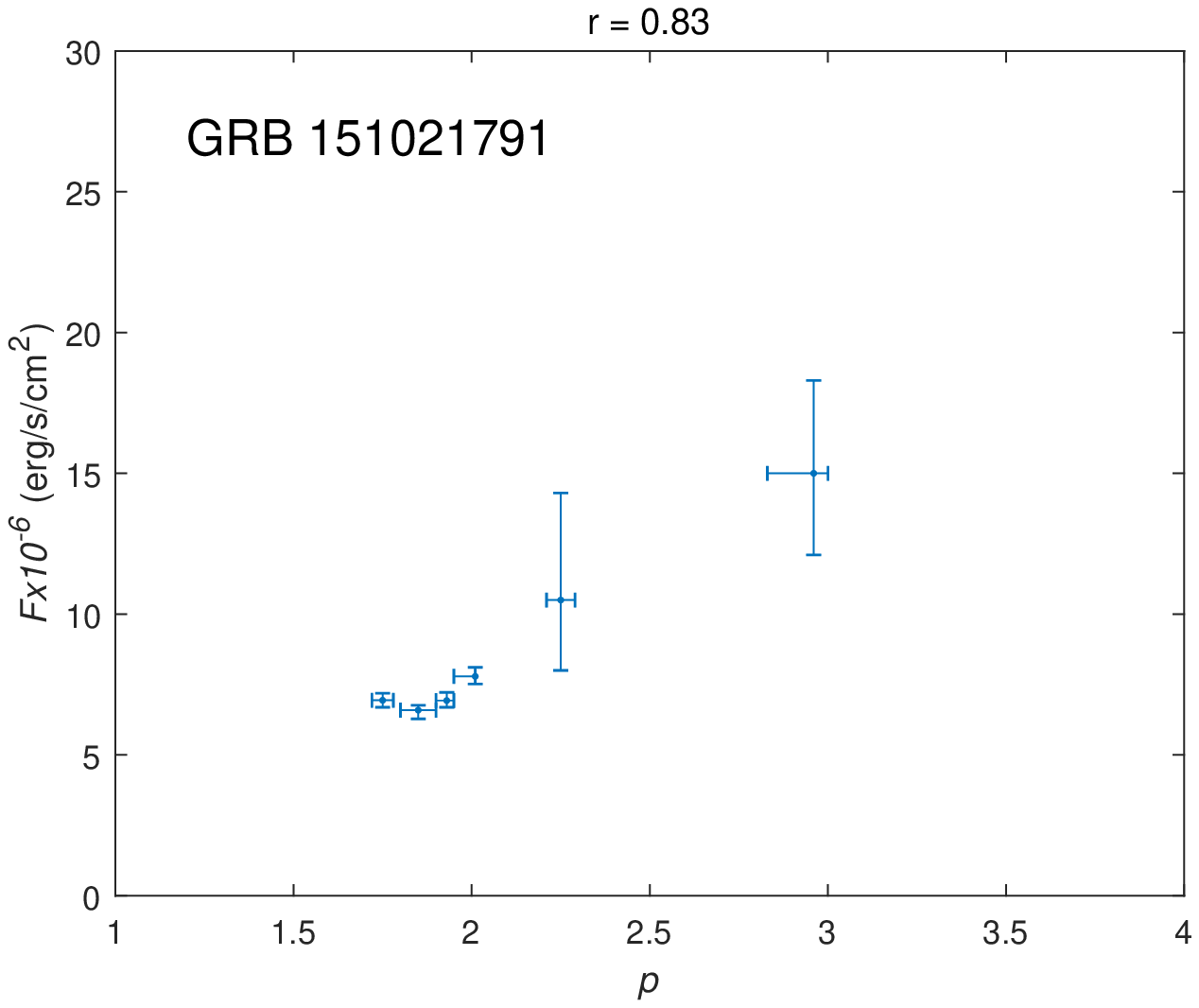}
 \includegraphics[width=5.8cm,height=4.2cm]{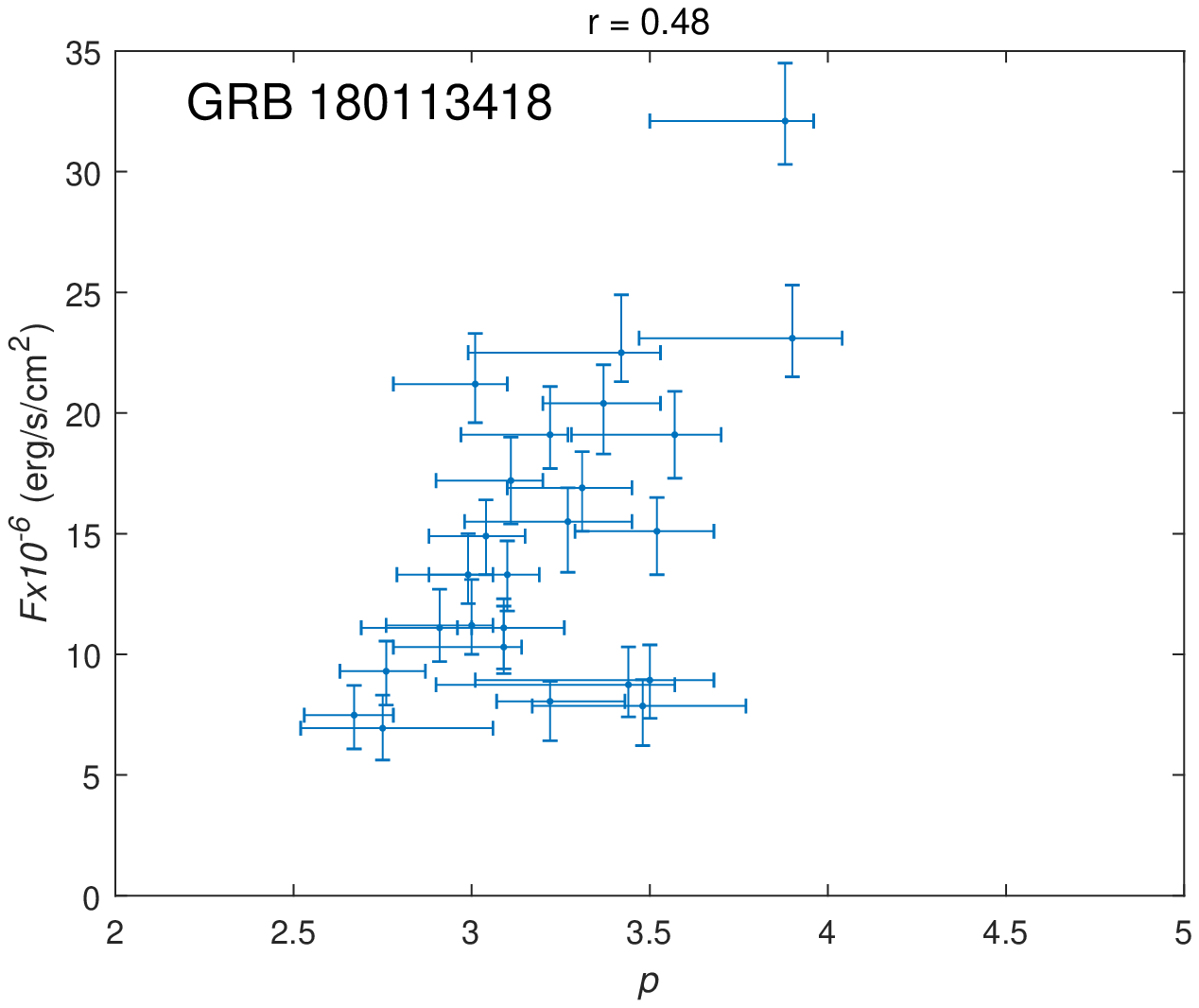}
 \caption{ Evolution of $p$ parameter in the SYNPL model (blue data points, right axis) of six sample bursts with the light curve (black curve, left axis) and correlations between $p$ and the flux}\label{p-evol}
\end{figure*}

The $F-p$ tracking behavior in Fig. \ref{p-evol} suggests that the electrons are distributed in a much narrower range as the flux approaches its peak. The slope of $p$ at the peak time of flux for some bursts is larger than $5$, which actually corresponds to the mono-energetic electron synchrotron spectrum with the electron energy and number concentrated in the minimal injection energy of electrons. The steep electron spectra is inconsistent with the typical slope expected in theoretical studies of the relativistic shock model (e.g., Kirk et al. 2000), where $p\sim2.3$. Even if we consider the synchrotron cooling, the typical theoretical index is $p\sim3.3$, which is still much smaller than is found in some bursts. This suggests that the electrons may be accelerated by other mechanisms.

\section*{Acknowledgements}
We acknowledge partial support by the Chinese Natural Science Foundation ( No. U1831135, 11763009, 11263006), Key Laboratory of Colleges and Universities in Yunnan Province for High-energy Astrophysics. B.B.Z acknowledges
 support by Fundamental Research Funds for the Central Universities
 (14380046), the National Key Research and Development Programs of China
 (2018YFA0404204), the National Natural Science Foundation of China (Grant
 Nos. 11833003, U2038105), the science research grants from the China Manned Space Project with NO.CMS-CSST-2021-B11, and the Program for Innovative Talents,
 Entrepreneur in Jiangsu. We acknowledge the use of public data from the
 Fermi Science Support Center (FSSC).

\begin{longrotatetable}
\begin{deluxetable*}{ccccccccccccccc}
\centering
\renewcommand\tabcolsep{0.95pt}
\renewcommand\arraystretch{1.2}
\tablecaption{Time-resolved spectral analysis results of GRB 130121835\label{chartable}}
\tablewidth{700pt}
\tabletypesize{\scriptsize}
\tablehead{
 % \begin{tabular}{ccccccccccccccc}
 % \hline\hline
 \colhead{$t_{start}$:$t_{end}$} &\colhead{Models} &\colhead{$\alpha/p$} &\colhead{E$_{p}$/$log(\gamma_{m})$/$log(\gamma_{th})$} &\colhead{PGSTAT/dof} &\colhead{BIC} &\colhead{F$\times10^{-6}$} &\colhead{Models} &\colhead{$\alpha$/$p_{1}$} &\colhead{$\beta$/p$_{2}$} &\colhead{$log(\gamma_{m})$} &\colhead{E$_{p}$/$log(\gamma_{b})$} &\colhead{PGSTAT/dof} &\colhead{BIC} &\colhead{F$\times10^{-6}$} \\
 \colhead{(1)} &\colhead{(2)} &\colhead{(3)} & \colhead{(4)} &\colhead{(5)} &\colhead{(6)} &\colhead{(7)} &\colhead{(8)} &\colhead{(9)} &\colhead{(10)} &\colhead{(11)} &\colhead{(12)} &\colhead{(13)} &\colhead{(14)} &\colhead{(15)}
 }
 \startdata
 % \hline
 0.00:0.87 & CPL & -1.11$_{-0.04}^{+0.08}$ & 2990.84$_{-410.86}^{+9.05}$ & 282.4/349 & 300.01 & 6.39$_{-0.99}^{+1.20}$ & BAND & -1.10$_{-0.05}^{+0.07}$ & -6.26$_{-11.01}^{+0.79}$ &$-$ & 2999.99$_{-418.31}^{+0.00}$ & 282.4/348 & 305.84 & 6.43$_{-1.05}^{+1.10}$ \\
 & SYNPL & 1.88$_{-0.07}^{+23.4}$ & 2.97$_{-0.06}^{+0.98}$ & 308.8/349 & 326.42 & 6.23$_{-1.58}^{+3.04}$ & SYNBPL & 1.13$_{-0.24}^{+0.17}$ & 28.41$_{-19.22}^{+1.59}$ & 2.45$_{-1.14}^{+0.22}$ & 4.28$_{-0.09}^{+0.07}$ & 283.7/347 & 313.03 & 7.09$_{-1.32}^{+1.73}$ \\
& SYNSE &$-$ & 3.96$_{-0.04}^{+0.02}$ &316.7/350 & 328.42 & 8.00$_{-1.63}^{+1.77}$
 & SYNFC &$-$ & 1.80$_{-0.05}^{+0.06}$ &1.69$_{-0.39}^{+1.20}$ &$-$ & 314.9/349 & 332.45 & 6.59$_{-0.86}^{+0.81}$ \\
 & SYNMAX &$-$ & 3.37$_{-0.06}^{+0.04}$ &302.1/350 & 313.87 & 6.77$_{-1.47}^{+1.45}$ \\
 0.87:1.73 & CPL & -1.15$_{-0.03}^{+0.05}$ & 2985.91$_{-635.28}^{+13.75}$ & 300.8/349 & 318.37 & 8.41$_{-1.50}^{+0.99}$ & BAND & -1.15$_{-0.03}^{+0.05}$ & -4.86$_{-12.55}^{+2.79}$ &$-$ & 2981.22$_{-632.11}^{+18.72}$ & 300.8/348 & 324.22 & 8.30$_{-1.38}^{+1.00}$ \\
 & SYNPL & 2.09$_{-0.11}^{+0.13}$ & 3.16$_{-0.07}^{+0.03}$ & 328.0/349 & 345.61 & 7.70$_{-1.33}^{+1.38}$ & SYNBPL & 1.10$_{-0.16}^{+0.18}$ & 29.04$_{-20.17}^{+0.59}$ & 1.65$_{-0.32}^{+0.98}$ & 4.21$_{-0.10}^{+0.09}$ & 303.8/347 & 333.14 & 8.53$_{-1.78}^{+1.78}$ \\
& SYNSE &$-$ & 3.77$_{-0.04}^{+0.04}$ & 356.5/350 & 368.22 & 5.05$_{-1.13}^{+1.55}$
 & SYNFC &$-$ & 1.81$_{-0.04}^{+0.05}$ &1.67$_{-0.35}^{+1.04}$ &$-$ & 348.6/349 & 366.21 & 9.05$_{-0.90}^{+0.87}$ \\
 & SYNMAX &$-$ & 3.26$_{-0.04}^{+0.04}$ &323.4/350 & 335.11 & 6.72$_{-1.15}^{+1.35}$ \\
 1.73:2.60 & CPL & -1.07$_{-0.03}^{+0.04}$ & 2602.41$_{-345.82}^{+213.82}$ & 316.3/349 & 333.86 & 11.50$_{-1.60}^{+1.50}$ & BAND & -1.07$_{-0.03}^{+0.04}$ & -8.42$_{-8.78}^{+2.24}$ &$-$ & 2641.91$_{-387.89}^{+180.26}$ & 316.3/348 & 339.75 & 11.60$_{-1.80}^{+1.40}$ \\
 & SYNPL & 4.66$_{-2.17}^{+12.92}$ & 3.65$_{-0.01}^{+0.15}$ & 368.7/349 & 386.33 & 8.44$_{-1.71}^{+2.36}$ & SYNBPL & 0.80$_{-0.27}^{+0.10}$ & 26.66$_{-15.42}^{+0.03}$ & 1.97$_{-0.67}^{+0.63}$ & 4.14$_{-0.08}^{+0.02}$ & 322.9/347 & 352.20 & 11.40$_{-1.90}^{+1.60}$ \\
& SYNSE &$-$ & 3.83$_{-0.03}^{+0.02}$ &376.1/350 & 387.81 & 8.82$_{-1.45}^{+1.65}$
 & SYNFC &$-$ & 1.80$_{-0.03}^{+0.03}$ &1.43$_{-0.14}^{+1.15}$ &$-$ & 459.9/349 & 477.53 & 12.10$_{-0.90}^{+0.90}$ \\
 & SYNMAX &$-$ & 3.29$_{-0.03}^{+0.03}$ &339.1/350 & 350.80 & 10.00$_{-1.20}^{+1.30}$ \\
 2.60:3.47 & CPL & -1.09$_{-0.04}^{+0.03}$ & 1432.36$_{-175.05}^{+260.78}$ & 346.9/349 & 364.50 & 11.80$_{-1.70}^{+2.10}$ & BAND & -1.09$_{-0.03}^{+0.04}$ & -9.90$_{-7.49}^{+3.75}$ &$-$ & 1452.60$_{-195.33}^{+230.69}$ & 346.9/348 & 370.37 & 11.90$_{-1.90}^{+2.00}$ \\
 & SYNPL & 3.30$_{-0.22}^{+0.31}$ & 3.43$_{-0.02}^{+0.04}$ & 367.2/349 & 384.78 & 10.70$_{-1.30}^{+1.80}$ & SYNBPL & 0.90$_{-0.39}^{+0.07}$ & 28.96$_{-18.38}^{+1.02}$ & 3.04$_{-0.67}^{+0.03}$ & 4.02$_{-0.09}^{+0.01}$ & 342.1/347 & 371.42 & 12.10$_{-1.80}^{+2.20}$ \\
& SYNSE &$-$ & 3.65$_{-0.01}^{+0.01}$ & 402.4/350 & 413.75 & 7.27$_{-0.54}^{+0.66}$
 & SYNFC &$-$ & 5.83$_{-1.93}^{+0.17}$ &4.43$_{-0.06}^{+0.09}$ & $-$ & 632.6/349 & 650.15 & 12.80$_{-1.00}^{+0.90}$ \\
 & SYNMAX &$-$ & 3.16$_{-0.02}^{+0.02}$ &347.8/350 & 359.55 & 11.10$_{-0.90}^{+1.00}$ \\
 3.47:4.33 & CPL & -1.12$_{-0.03}^{+0.04}$ & 1139.18$_{-184.47}^{+218.03}$ & 333.3/349 & 350.93 & 9.88$_{-1.63}^{+1.92}$ & BAND & -1.12$_{-0.04}^{+0.04}$ & -8.22$_{-9.28}^{+2.12}$ &$-$ & 1138.01$_{-177.37}^{+209.73}$ & 333.4/348 & 356.81 & 9.92$_{-1.78}^{+1.81}$ \\
 & SYNPL & 3.22$_{-0.15}^{+0.36}$ & 3.38$_{-0.03}^{+0.03}$ & 344.1/349 & 361.72 & 9.75$_{-1.53}^{+1.34}$ & SYNBPL & 1.49$_{-0.73}^{+0.19}$ & 28.82$_{-18.33}^{+1.18}$ & 3.16$_{-1.03}^{+0.05}$ & 4.03$_{-0.12}^{+0.03}$ & 320.5/347 & 349.82 & 10.70$_{-2.30}^{+2.30}$ \\
& SYNSE &$-$ & 3.59$_{-0.01}^{+0.01}$ & 383.2/350 & 394.90 & 6.08$_{-0.42}^{+0.44}$
 & SYNFC &$-$ & 5.93$_{-1.52}^{+0.07}$ &4.31$_{-0.07}^{+0.04}$ &$-$ & 557.9/349 & 575.49 & 12.00$_{-1.20}^{+1.00}$ \\
 & SYNMAX &$-$ & 3.10$_{-0.02}^{+0.02}$ &327.5/350 & 339.18 & 9.19$_{-0.70}^{+0.81}$ \\
 4.33:5.20 & CPL & -1.23$_{-0.03}^{+0.03}$ & 1738.74$_{-280.67}^{+401.92}$ & 363.7/349 & 381.33 & 9.50$_{-1.81}^{+1.84}$ & BAND & -1.24$_{-0.03}^{+0.03}$ & -15.40$_{-1.92}^{+9.38}$ &$-$ & 1727.71$_{-279.43}^{+429.73}$ & 363.7/348 & 387.19 & 9.35$_{-1.73}^{+1.91}$ \\
 & SYNPL & 2.51$_{-0.00}^{+0.28}$ & 3.22$_{-0.00}^{+0.06}$ & 366.1/349 & 383.68 & 9.59$_{-1.93}^{+0.79}$ & SYNBPL & 1.78$_{-0.43}^{+0.19}$ & 27.77$_{-18.17}^{+2.21}$ & 3.09$_{-0.16}^{+0.06}$ & 4.15$_{-0.14}^{+0.04}$ & 352.5/347 & 381.81 & 9.41$_{-2.11}^{+1.97}$ \\
& SYNSE &$-$ & 3.54$_{-0.01}^{+0.02}$ & 430.1/350 & 441.87 & 4.14$_{-0.32}^{+0.36}$
 & SYNFC &$-$ & 5.95$_{-2.33}^{+0.04}$ &4.38$_{-0.13}^{+0.07}$ &$-$ & 463.2/349 & 480.77 & 10.20$_{-1.10}^{+1.00}$ \\
 & SYNMAX &$-$ & 3.06$_{-0.02}^{+0.02}$ &372.1/350 & 383.79 & 6.45$_{-0.58}^{+0.72}$ \\
 5.20:6.07 & CPL & -1.30$_{-0.02}^{+0.03}$ & 2300.42$_{-414.19}^{+397.88}$ & 382.7/349 & 400.34 & 11.30$_{-2.0}^{+1.70}$ & BAND & -1.29$_{-0.03}^{+0.02}$ & -17.34$_{-0.08}^{+11.77}$ &$-$ & 2215.80$_{-346.82}^{+469.93}$ & 382.8/348 & 406.24 & 11.10$_{-1.70}^{+1.90}$ \\
 & SYNPL & 2.64$_{-0.16}^{+0.06}$ & 3.21$_{-0.04}^{+0.02}$ & 371.9/349 & 389.52 & 9.33$_{-0.89}^{+1.72}$ & SYNBPL & 2.07$_{-0.19}^{+0.17}$ & 26.80$_{-17.44}^{+3.20}$ & 3.12$_{-0.06}^{+0.04}$ & 4.29$_{-0.12}^{+0.07}$ & 359.9/347 & 389.24 & 10.60$_{-1.90}^{+1.70}$ \\
& SYNSE &$-$ & 3.51$_{-0.01}^{+0.02}$ & 461.6/350 & 473.35 & 4.10$_{-0.29}^{+0.33}$
 & SYNFC &$-$ & 5.99$_{-1.99}^{+0.01}$ &4.38$_{-0.11}^{+0.04}$ &$-$ & 460.3/349 & 477.94 & 11.20$_{-1.10}^{+0.90}$ \\
 & SYNMAX &$-$ & 3.02$_{-0.04}^{+0.02}$ &395.2/350 & 406.94 & 6.35$_{-0.55}^{+0.67}$ \\
 6.07:6.93 & CPL & -1.26$_{-0.03}^{+0.03}$ & 1627.65$_{-246.78}^{+359.52}$ & 359.2/349 & 376.77 & 10.40$_{-1.70}^{+1.90}$ & BAND & -0.83$_{-0.46}^{+0.08}$ & -1.77$_{-15.47}^{+0.07}$ &$-$ & 256.87$_{-36.81}^{+1744.40}$ & 349.8/348 & 373.24 & 10.70$_{-2.40}^{+1.60}$ \\
 & SYNPL & 2.66$_{-0.12}^{+0.12}$ & 3.22$_{-0.03}^{+0.03}$ & 346.0/349 & 363.63 & 9.95$_{-1.16}^{+1.55}$ & SYNBPL & 1.95$_{-0.14}^{+0.25}$ & 28.43$_{-19.20}^{+1.56}$ & 3.10$_{-0.03}^{+0.06}$ & 4.17$_{-0.10}^{+0.06}$ & 332.0/347 & 361.34 & 10.40$_{-1.90}^{+1.60}$ \\
& SYNSE &$-$ & 3.52$_{-0.01}^{+0.01}$ & 437.4/350 & 449.15 &4.52$_{-0.30}^{+0.32}$
 & SYNFC &$-$ & 5.88$_{-1.95}^{+0.12}$ &4.31$_{-0.06}^{+0.08}$ &$-$ & 463.4/349 & 481.01 & 11.50$_{-1.00}^{+1.20}$ \\
 & SYNMAX &$-$ & 3.02$_{-0.02}^{+0.02}$ &362.1/350 & 373.78 & 6.95$_{-0.58}^{+0.63}$ \\
 6.93:7.80 & CPL & -1.16$_{-0.03}^{+0.04}$ & 1093.38$_{-172.04}^{+206.64}$ & 374.9/349 & 392.48 & 11.70$_{-1.90}^{+2.0}$ & BAND & -0.98$_{-0.20}^{+0.09}$ & -2.11$_{-14.76}^{+0.14}$ &$-$ & 481.93$_{-93.95}^{+816.11}$ & 371.3/348 & 394.72 & 11.70$_{-1.90}^{+2.00}$ \\
 & SYNPL & 3.09$_{-0.15}^{+0.17}$ & 3.33$_{-0.03}^{+0.02}$ & 367.6/349 & 385.15 & 12.10$_{-1.70}^{+1.20}$ & SYNBPL & 1.85$_{-0.20}^{+0.29}$ & 19.41$_{-8.59}^{+8.70}$ & 3.18$_{-0.04}^{+0.06}$ & 4.07$_{-0.07}^{+0.06}$ & 342.3/347 & 371.57 & 12.80$_{-1.90}^{+2.00}$ \\
& SYNSE &$-$ & 3.56$_{-0.01}^{+0.01}$ & 438.6/350 & 450.35 & 6.90$_{-0.38}^{+0.42}$
 & SYNFC &$-$ & 6.00$_{-1.35}^{+0.00}$ &4.30$_{-0.07}^{+0.03}$ &$-$ & 624.2/349 & 641.82 & 14.80$_{-1.30}^{+1.10}$ \\
 & SYNMAX &$-$ & 3.07$_{-0.01}^{+0.02}$ &358.1/350 & 369.81 & 10.40$_{-0.70}^{+0.80}$ \\
 7.80:8.67 & CPL & -1.38$_{-0.03}^{+0.04}$ & 1517.33$_{-426.4}^{+634.99}$ & 324.0/349 & 341.61 & 8.01$_{-2.34}^{+2.14}$ & BAND & -1.05$_{-0.33}^{+0.02}$ & -1.88$_{-9.79}^{+0.03}$ &$-$ & 245.76$_{-7.99}^{+1374.40}$ & 309.3/348 & 332.80 & 7.51$_{-2.32}^{+2.61}$ \\
 & SYNPL & 2.76$_{-0.14}^{+0.12}$ & 3.15$_{-0.03}^{+0.02}$ & 311.3/349 & 328.94 & 7.46$_{-1.01}^{+1.42}$ & SYNBPL & 2.28$_{-0.15}^{+0.29}$ & 17.23$_{-8.59}^{+8.70}$ & 3.06$_{-0.04}^{+0.07}$ & 4.22$_{-0.12}^{+0.19}$ & 304.0/347 & 333.31 & 7.93$_{-1.69}^{+1.84}$ \\
& SYNSE &$-$ & 3.44$_{-0.01}^{+0.01}$ & 382.5/350 & 394.24 & 3.32$_{-0.20}^{+0.22}$
 & SYNFC &$-$ & 5.94$_{-1.77}^{+0.06}$ &4.20$_{-0.10}^{+0.03}$ &$-$ & 348.0/349 & 365.63 & 9.26$_{-1.18}^{+1.18}$ \\
 & SYNMAX &$-$ & 2.92$_{-0.02}^{+0.02}$ &327.2/350 & 338.93 & 4.71$_{-0.37}^{+0.39}$ \\
 8.67:9.53 & CPL & -1.42$_{-0.02}^{+0.03}$ & 2962.80$_{-821.9}^{+36.41}$ & 377.8/349 & 395.40 & 9.21$_{-1.45}^{+0.78}$ & BAND & -1.04$_{-0.39}^{+0.00}$ & -1.76$_{-15.39}^{+0.00}$ &$-$ & 215.02$_{-0.00}^{+2659.47}$ & 373.2/348 & 396.61 & 7.80$_{-0.24}^{+2.18}$ \\
 & SYNPL & 2.48$_{-0.01}^{+0.10}$ & 3.07$_{-0.04}^{+0.03}$ & 372.3/349 & 389.86 & 7.92$_{-1.16}^{+1.28}$ & SYNBPL& 2.17$_{-0.13}^{+0.21}$ & 26.35$_{-17.94}^{+3.63}$ & 3.00$_{-0.06}^{+0.03}$ & 4.33$_{-0.11}^{+0.19}$ & 365.4/347 & 394.70 & 8.22$_{-1.69}^{+1.79}$ \\
& SYNSE &$-$ & 3.41$_{-0.01}^{+0.02}$ & 478.3/350 & 490.04 & 2.70$_{-0.17}^{+0.20}$
 & SYNFC &$-$ & 5.97$_{-2.07}^{+0.03}$ &4.28$_{-0.12}^{+0.04}$ &$-$ & 390.7/349 & 408.28 & 9.03$_{-1.19}^{+1.10}$ \\
 & SYNMAX &$-$ & 2.90$_{-0.02}^{+0.02}$ &414.8/350 & 426.48 & 3.98$_{-0.34}^{+0.37}$ \\
 9.53:10.40 & CPL & -1.47$_{-0.03}^{+0.04}$ & 2976.46$_{-1036.1}^{+22.16}$ & 354.7/349 & 372.27 & 6.48$_{-1.37}^{+0.71}$ & BAND & -1.47$_{-0.03}^{+0.04}$ & -9.57$_{-7.54}^{+4.32}$ &$-$ & 2880.26$_{-916.34}^{+119.37}$ & 354.7/348 & 378.18 & 6.40$_{-1.26}^{+0.68}$ \\
 & SYNPL & 2.48$_{-0.10}^{+0.12}$ & 2.99$_{-0.03}^{+0.05}$ & 351.4/349 & 368.96 & 5.70$_{-1.80}^{+1.10}$ & SYNBPL & 2.29$_{-0.24}^{+0.19}$ & 20.82$_{-13.54}^{+5.07}$ & 2.96$_{-0.12}^{+0.05}$ & 4.34$_{-0.15}^{+0.76}$ & 347.6/347 & 376.89 & 5.63$_{-1.42}^{+1.58}$ \\
& SYNSE &$-$ & 3.37$_{-0.02}^{+0.02}$ & 430.9/350 & 442.66 & 1.85$_{-0.14}^{+0.15}$
 & SYNFC &$-$ & 5.98$_{-2.28}^{+0.02}$ &4.20$_{-0.15}^{+0.06}$ &$-$ & 356.8/349 & 374.35 & 6.41$_{-1.08}^{+1.10}$ \\
 & SYNMAX &$-$ & 2.85$_{-0.02}^{+0.03}$ &384.2/350 & 395.93 & 2.63$_{-0.26}^{+0.28}$ \\
 10.40:11.27 & CPL & -1.41$_{-0.03}^{+0.04}$ & 2983.48$_{-583.05}^{+15.89}$ & 355.9/349 & 373.52 & 5.87$_{-0.87}^{+0.79}$ & BAND & -1.41$_{-0.04}^{+0.04}$ & -14.53$_{-2.53}^{+9.99}$ &$-$ & 2998.47$_{-1.45}^{+599.58}$ & 355.9/348 & 379.37 & 5.96$_{-0.95}^{+1.71}$ \\
 & SYNPL & 2.09$_{-0.1}^{+0.01}$ & 2.78$_{-1.51}^{+0.12}$ & 356.0/349 & 373.58 & 6.64$_{-0.64}^{+1.61}$ & SYNBPL & 1.90$_{-0.07}^{+0.15}$ & 16.74$_{-12.00}^{+8.54}$ & 2.63$_{-1.34}^{+0.27}$ & 4.50$_{-0.18}^{+0.80}$ & 353.2/347 & 382.48 & 7.18$_{-1.66}^{+1.23}$ \\
& SYNSE &$-$ & 3.44$_{-0.03}^{+0.02}$ & 476.2/350 & 487.93 & 1.70$_{-0.19}^{+0.23}$
 & SYNFC &$-$ & 3.10$_{-0.73}^{+2.13}$ &4.39$_{-0.06}^{+0.81}$ &$-$ & 355.1/349 & 372.67 & 6.81$_{-1.05}^{+0.87}$ \\
 & SYNMAX &$-$ & 2.92$_{-0.03}^{+0.04}$ &420.4/350 & 432.17 & 2.50$_{-0.35}^{+0.44}$ \\
 11.27:12.13 & CPL & -1.40$_{-0.03}^{+0.05}$ & 2993.46$_{-761.98}^{+6.24}$ & 405.4/349 & 422.96 & 5.95$_{-1.05}^{+0.78}$ & BAND & -1.39$_{-0.03}^{+0.05}$ & -8.77$_{-8.27}^{+3.57}$ &$-$ & 2986.05$_{-762.99}^{+13.90}$ & 405.4/348 & 428.86 & 6.01$_{-1.13}^{+0.78}$ \\
 & SYNPL & 2.24$_{-0.16}^{+0.05}$ & 2.96$_{-0.21}^{+0.03}$ & 417.8/349 & 435.42 & 5.87$_{-1.12}^{+1.36}$ & SYNBPL & 1.80$_{-0.11}^{+0.09}$ & 25.52$_{-16.89}^{+0.50}$ & 2.75$_{-1.46}^{+0.17}$ & 4.27$_{-0.10}^{+0.10}$ & 406.6/347 & 435.92 & 6.43$_{-1.05}^{+1.10}$ \\
& SYNSE &$-$ & 3.42$_{-0.03}^{+0.02}$ & 504.6/350 & 516.33 & 1.60$_{-0.18}^{+0.21}$
 & SYNFC &$-$ & 6.00$_{-2.65}^{+0.00}$ &4.34$_{-0.17}^{+0.11}$ &$-$ & 413.9/349 & 431.49 & 5.89$_{-0.99}^{+0.90}$ \\
 & SYNMAX &$-$ & 2.93$_{-0.03}^{+0.04}$ &456.2/350 & 467.93 & 2.56$_{-0.36}^{+0.46}$ \\
 12.13:13.00 & CPL & -1.46$_{-0.04}^{+0.05}$ & 2600.74$_{-750.94}^{+197.62}$ & 310.8/349 & 328.40 & 4.81$_{-0.98}^{+0.81}$ & BAND & -1.46$_{-0.04}^{+0.04}$ & -17.56$_{-2.44}^{+11.93}$ &$-$ & 2537.86$_{-686.17}^{+247.97}$ & 310.8/348 & 334.27 & 4.79$_{-0.99}^{+0.82}$ \\
 & SYNPL & 2.30$_{-0.16}^{+0.04}$ & 2.88$_{-1.26}^{+0.04}$ & 324.6/349 & 342.24 & 4.78$_{-1.08}^{+1.41}$ & SYNBPL & 1.93$_{-0.13}^{+0.09}$ & 21.11$_{-12.08}^{+5.10}$ & 2.62$_{-1.34}^{+0.22}$ & 4.19$_{-0.10}^{+0.10}$ & 312.4/347 & 341.68 & 4.57$_{-1.00}^{+1.38}$ \\
& SYNSE &$-$ & 3.39$_{-0.02}^{+0.03}$ & 402.8/350 & 414.50 & 1.42$_{-0.15}^{+0.17}$
 & SYNFC &$-$ & 5.86$_{-2.03}^{+0.14}$ & 4.14$_{-0.10}^{+0.11}$ &$-$ & 314.4/349 & 332.03 & 4.53$_{-0.79}^{+1.18}$ \\
 & SYNMAX &$-$ & 2.87$_{-0.03}^{+0.04}$ &355.7/350 & 367.47 & 2.05$_{-0.26}^{+0.34}$ \\
 % \hline
\enddata
\tablecomments{We list the start and stop times of each time slice of each burst (Col. 1), fitting models (Cols. 2 and 8), model fitting parameters (Cols. 3 $\sim$ 4 and Cols. 9 $\sim$ 12), PGSTAT/dof (Cols. 5 and 13), Bayesian Information Criteria (BIC) (Cols. 6 and 14) and energy flux (F) in 1keV $\sim10^{4}$ keV (Cols. 7 and 15). All time parameters have units of s, energies have units of keV, and fluxes have units of erg $s^{-1} cm^{-2}$.}
\end{deluxetable*}
\end{longrotatetable}

\begin{table*}
\centering
\renewcommand\tabcolsep{2.2pt}
\renewcommand\arraystretch{1.4}
\caption{Values of the median and standard deviation of the parameter distributions.}
\begin{tabular}{cccccccccc} % four columns, alignment for each
\hline\hline
 Models & $\alpha/p/p_{1}$ & $E_{p}/log(\gamma_{b})$ & $\beta$ & $log(\gamma_{m})/log(\gamma_{th})$ \\
\hline
 CPL &-0.92$\pm0.45$ &328.63$\pm69.98$ &$-$ &$-$\\
 BAND &-0.82$\pm0.52$ &266.48$\pm64.04$ &-2.88$\pm2.45$ &$-$\\
 SYNPL &4.63$\pm2.01$ &$-$ &$-$ &3.34$\pm0.40$\\
 SYNSE &$-$ &$-$ &$-$ &3.46$\pm0.22$\\
 SYNBPL &2.59$\pm2.72$ &3.74$\pm0.53$ &$-$ &3.20$\pm0.72$\\
 SYNMAX &$-$ &$-$ &$-$ &2.94$\pm0.27$\\
\hline
\end{tabular}
\label{fit_data}
\end{table*}

\end{document}